\documentclass[preprint]{aastex63}
\received{}
\revised{}
\accepted{}
\newcommand{\jjj}{BDR\,J1750+3809}
\newcommand{\ch}{CH$_4$s }
\newcommand{\sa}{2MASS\,J17500008+3809276}
\usepackage{lineno}
\submitjournal{ApJL}
\shorttitle{Radio selected BD}
\shortauthors{Vedantham et al.}
\begin{document}

\title{Direct radio discovery of a cold brown dwarf}

\correspondingauthor{H. K. Vedantham}
\email{vedantham@astron.nl}
\author{H. K. Vedantham}
\affiliation{ASTRON, Netherlands Institute for Radio Astronomy, Oude Hoogeveensedijk 4, 7991PD, Dwingeloo, The Netherlands}
\affiliation{Kapteyn Astronomical Institute, University of Groningen, The Netherlands}
\author{J. R. Callingham}
\affiliation{Leiden Observatory, Leiden University, PO Box 9513, 2300 RA Leiden, The Netherlands}
\affiliation{ASTRON, Netherlands Institute for Radio Astronomy, Oude Hoogeveensedijk 4, 7991PD, Dwingeloo, The Netherlands}
\author{T. W. Shimwell}
\affiliation{ASTRON, Netherlands Institute for Radio Astronomy, Oude Hoogeveensedijk 4, 7991PD, Dwingeloo, The Netherlands}
\affiliation{Leiden Observatory, Leiden University, PO Box 9513, 2300 RA Leiden, The Netherlands}
\author{T. Dupuy}
\affiliation{Institute for Astronomy, University of Edinburgh, Blackford Hill, Edinburgh, EH9 3HJ,  United Kingdom}
\affiliation{Gemini Observatory, Northern Operations Center, 670 N. A'ohoku Place, Hilo, HI 96720, USA}
\author{William M. J. Best}
\affiliation{University of Texas at Austin, Department of Astronomy, 2515 Speedway C1400, Austin, TX 78712, USA}
\affiliation{Visiting Astronomer at the Infrared Telescope Facility, which is operated by the University of Hawaii under contract 80HGTR19D0030 with the National Aeronautics and Space Administration.}
\author{Michael C. Liu}
\affiliation{Visiting Astronomer at the Infrared Telescope Facility, which is operated by the University of Hawaii under contract 80HGTR19D0030 with the National Aeronautics and Space Administration.}
\affiliation{Institute for Astronomy, University of Hawai`i, 2680 Woodlawn Drive, Honolulu, HI 96822}
\author{Zhoujian Zhang}
\affiliation{Institute for Astronomy, University of Hawai`i, 2680
  Woodlawn Drive, Honolulu, HI 96822}
\author{K. De}
\affiliation{Cahill center for astronomy and astrophysics, California Institute of Technology, 1200 E California Blvd. Pasadena, CA, 91125}
\author{L. Lamy}
\affiliation{LESIA, CNRS – Observatoire de Paris, PSL 92190, Meudon, France}
\author{P. Zarka}
\affiliation{LESIA, CNRS – Observatoire de Paris, PSL 92190, Meudon, France}
\author{H.~J.~A.~R\"{o}ttgering}
\affiliation{Leiden Observatory, Leiden University, PO Box 9513, 2300 RA Leiden, The Netherlands}
\author{A. Shulevski}
\affiliation{Leiden Observatory, Leiden University, PO Box 9513, 2300 RA Leiden, The Netherlands}

\begin{abstract}
Magnetospheric processes seen in gas-giants such as aurorae and circularly-polarized cyclotron maser radio emission have been detected from some brown dwarfs. However, previous radio observations targeted known brown dwarfs discovered via their infrared emission. Here we report the discovery of \jjj, a circularly polarized radio source detected around 144\,MHz with the LOFAR telescope. Follow-up near-infrared photometry and spectroscopy show that \jjj\ is a cold methane dwarf of spectral type T$6.5\pm 1$ at a distance of $65^{+9}_{-8}\,{\rm pc}$. The quasi-quiescent radio spectral luminosity of \jjj\,is $\approx 5\times 10^{15}\,{\rm erg}\,{\rm s}^{-1}\,{\rm Hz}^{-1}$ which is over two orders of magnitude larger than that of the known population of comparable spectral type. This could be due to a preferential geometric alignment or an electrodynamic interaction with a close companion. In addition, as the emission is expected to occur close to the electron gyro-frequency, the magnetic field strength at the emitter site in \jjj\,is $B\gtrsim 25\,{\rm G}$, which is comparable to planetary-scale magnetic fields. Our discovery suggests that low-frequency radio surveys can be employed to discover sub-stellar objects that are too cold to be detected in infrared surveys.
\end{abstract}

\section{Introduction} \label{sec:intro}
The generation and dissipation of magnetic flux in stars and planets is pivotal in driving violent stellar activity and determining the space plasma environment around exoplanets respectively \citep{schrijver2008,schwenn2006}. On cool objects where Zeeman splitting observations are difficult (later than type M typically), observation of cyclotron emission, that falls in the radio band, is the only known technique to directly measure the strength and topology of the objects' magnetic fields. 

Brown dwarfs (BD), with masses between that of stars and planets, display optical aurorae \citep{hallinan2015} and the associated auroral radio emission \citep{hallinan2015,kao2018,pineda2017,nichols2012} powered by the electron cyclotron maser instability \citep{hallinan2008,wu,treumann}.  In addition, because there appears to be no clear demarcation between the atmospheres and magnetospheres of the smallest coldest brown dwarfs and the largest planets \citep{exoplanet-handbook}, radio observations at the end of the BD sequence are expected to provide a tantalizing glimpse into magnetospheric properties of exoplanets \citep{kao2018,kao2020}. 

\citet{christensen} have argued that the magnetic fields of planets, brown dwarfs, and low-mass stars of sufficiently rapid rotation are dipolar and that the field strength scales with the heat-flux from the bodies' interior. The simplicity and universality of this law is a giant leap in modeling exoplanet atmospheres and habitability. The law can be tested at the low-mass end by measuring the magnetic fields of a sample of cold brown dwarfs and exoplanets via radio observations of their cyclotron emission \citep{kao2016,kao2018}.\footnote{The emission happens at the cyclotron frequency, $\nu_{\rm c}\approx 2.8(B/{\rm Gauss})\,{\rm MHz}$ or its second harmonic \citep{melrose1982}, where $B$ is the magnetic field strength.} 

Since their discovery as  radio-emitters \citep{berger2001}, radio surveys of known BDs have primarily been carried out at gigahertz frequencies that can only detect cyclotron emission from objects with kG-level magnetic fields \citep[see the compilation of ][Chapter 28]{pineda2017, williams2018}. Observations at much lower frequencies that probe `planetary scale' magnetic fields (few to tens of Gauss) are necessary to test the scaling law in the exoplanet-regime. Low frequency observations are now being carried out thanks to the advent of sensitive metre-wave telescopes such as LOFAR \citep{lofar}, and the wide-area surveys they facilitate such as the LoTSS survey \citep{lotss1,lotss2}. Low-frequency searches have so far been unsuccessful  \citep{bastian2000,lazio2004,hallinan2013,burningham2016,lynch2017,lenc2018}.

Searching for circularly polarized radio sources has proved to be a powerful technique to identify coherent stellar radio emission \citep{lynch,gj1151,joe-pop}. There are three known types of radio emitter with high circularly polarized (CP) fraction: (a) stars, (b) brown dwarfs and planets, and (c) pulsars. Lack of an optical counterpart to a CP source generally rules out a stellar association. We are currently following up such sources in the LoTSS survey \citep{lotss1,lotss2} data with NIR photometry and radio pulsation search to distinguish between the remaining two options. Here we report our first discovery from this effort--- \jjj. We will leave the overall counts and population statistics of unassociated CP sources for future work, save mention that \jjj\, stood out due to its high CP fraction (see \S 2.1 below) and that follow-up near-infrared photometric observations show the object to be a cold brown dwarf (see \S 2.2 below).

\jjj\, is the first radio-selected substellar object, which demonstrates that such objects can be directly discovered in a sensitive wide-area radio surveys. Because, the intensity of magnetospheric radio emission, that is non-thermal in nature, need not have a one-to-one scaling with the object's infrared luminosity, which is thermal in nature, \jjj's discovery also shows that ongoing low-frequency radio surveys could discover objects that are too cold and/or distant to be discovered and studied via their infrared emission.

\section{Discovery and follow-up}
\subsection{Radio properties}
\jjj\, was discovered as a radio source in an 8-hr LOFAR exposure between 120 and 167\,MHz with a high average CP fraction of $\approx 96^{+4}_{-20}\%$. The field containing \jjj\, was covered by two partially-overlapping LoTSS survey pointings, which were observed approximately six months apart in 2018. The radio source was only detected in one exposure (Fig. \ref{fig:radio_montage}). Separately, we obtained another LOFAR exposure centered on \jjj\, in January 2020. We re-detected the source in total intensity at low significance ($\approx 4\sigma$), but not in circular polarization. Forced CP photometry yields a polarized fraction of $12\pm16\%$. The radio source has not been detected previously, including in the first-epoch of the ongoing VLA Sky Survey at $2-4$\,GHz \citep[image noise of $\approx 0.1\,{\rm mJy}$; ][]{vlass}. Further details of radio data processing are given in Appendix A.

\subsection{Identification as a cold brown dwarf}
We searched publicly available optical and NIR archives for an association with the radio source. The source has no counterpart in the Pan-STARRS \citep{ps1}, 2MASS \citep{2mass} or AllWISE \citep{wise} survey catalogs. We found a faint ($6\sigma$ level) $J$-band detection (Fig. \ref{fig:nir_montage}) positionally coincident with \jjj\, in the UKIRT Hemisphere Survey \citep[UHSDR1; ][]{uhsdr1}.

To confirm the UKIRT detection and constrain the NIR colors, we obtained a $Ks$ band image of the source with the Wide-field Infrared Camera (WIRC; \citealt{Wilson2003}) on the Palomar 200-inch telescope. The data were reduced and stacked using a custom data reduction pipeline described in \citet{De2020}. With an effective exposure of 10\,min, we did not secure a detection at the location of the UKIRT J-band source (Fig. \ref{fig:nir_montage}). However, the $\approx 6$ year baseline between the two 200-inch and UKIRT exposures, and the unknown proper motion of \jjj, meant that we could not be certain if sub-threshold (low significance) detections in the $Ks$-band image could be associated. 

We obtained time for $J-$ and $Y$-band photometry on the Gemini-North telescope (program ID GN-2019B-DD-105). Because the workhorse imager, NIRI \citep{niri}, was unavailable at that time, we obtained imaging exposures through the acquisition keyhole of the GNIRS spectrometer \citep{gnirs}. This option yields a sensitivity comparable to that of NIRI but with a small field of view. The observing conditions did not permit the transfer of calibration solutions from photometric standards. We therefore tied our photometry to the nearby star \sa\, (Star A hereafter; see appendix for further details), which also fortuitously ensures a correction for any interstellar opacity effects.

We detected the counterpart to \jjj\, in both $J$ and $Y$ bands in the GNIRS keyhole images (Fig. \ref{fig:nir_montage}). 
We used the known position of the source from the 2019-10-19 GNIRS exposures to search for a sub-threshold detection in the $Ks$ band data from 2019-09-07. A forced photometric extraction yielded a faint $3\sigma$ detection.

We also found a $\approx 5\sigma$ detection in the $W1$ and $W2$ channels of the unWISE catalog \citep{unwise} that is an un-blurred co-addition of all available WISE exposures. Bulk of the WISE exposures of the field around \jjj,\, were taken in 2010. The WISE detections are consistent with the proper-motion corrected position of \jjj,\, (see \S2.4 below) within errors. The NIR colors ($Y-J$, $J-H$ and $J-W2$ for e.g.; see also Fig. \ref{fig:nir-color}) identify the object as a cold brown dwarf of spectral class T.

\subsection{Spectral type}
T-dwarfs are characterized by the presence of methane in their atmosphere \citep{fegley1996,kirkpatrick1999} due to their low surface temperatures that range from a few hundred to $\sim 1000\,{\rm K}$ \citep{nakajima2004}. To confirm the presence of atmospheric methane, we obtained further exposures using the NIRI instrument on the Gemini North telescope in $H$-band and the \ch band to perform `methane-imaging', which is a reliable technique for discovery and spectral typing of cool BDs \citep{rosenthal1996,tinney2005}. We detected \jjj\, in both filters at high significance (Fig. \ref{fig:nir_montage}). Based on the observed $H-$\ch colors of the object and the relationship of \citet[][their Eqn. 2 and Fig 4]{liu2008}, we estimate a spectral type of T$7.5\pm1.5$, confirming that \jjj\, is at the end of the T-dwarf sequence. 

Separately, we obtained a low-resolution ($R \approx$100) spectrum of \jjj\ on 2020~October~4~UT using the near-IR spectrograph SpeX \citep{2003PASP..115..362R} on NASA's Infrared Telescope Facility (IRTF) located on Maunakea, Hawaii.  Figure~\ref{fig:plot-spectra} shows the reduced spectrum of
\jjj, which has been flux-calibrated based on its J-band magnitude from UKIRT (Table~\ref{tab:phot-astro}). While the S/N is low ($\approx$6 per pixel in the $J$-band
peak), the spectrum clearly shows the strong water and methane
absorption bands that are the hallmarks of late-T dwarfs. We classified \jjj\ from the system of five spectral indices
established by \citet{2006ApJ...637.1067B}, resulting in a spectral type of T6.2$\pm$1.2. We also visually classified \jjj\ by comparing with IRTF/Spex spectra of the late-T standards from \citet{2006ApJ...637.1067B} and \citet{2011ApJ...743...50C}, finding a type of T7. Considering both the index and visual types, we adopt a final type of
T6.5$\pm$1.0.

All measurements of flux-density and position estimates are summarized in Table\,\ref{tab:phot-astro} for quick reference, while the Appendix provides further details of the observational setup and data processing.

\subsection{Distance and Proper motion}
We placed \jjj\ on the $J$ versus $J-W2$ color-magnitude relationship of cold methane dwarfs from \citet{leggett2017} (see Fig. \ref{fig:color-mag}) to find a distance of $d=70^{+25}_{-35}\,$pc. We estimated a more accurate photometric distance to \jjj\ using the spectral
type-absolute magnitude relation from \citet{dupuy2012}. For late-T dwarfs, the $W2$ band has the smallest intrinsic scatter to the relation ($\approx 0.19$\,mag), so we use this band, even though its observed
photometry has larger uncertainties than our near-IR photometry. We used a Monte Carlo calculation to account for the uncertainties in the spectral type (assumed to be uniformly distributed), the $W2$ photometry (normally distributed), and the relation's intrinsic scatter (normally distributed). The resulting distance modulus is $4.08\pm0.28$~mag, corresponding to a distance of $65_{-8}^{+9}$\,pc. (The same calculation
using $J$~band gives a consistent result, $55_{-10}^{+12}$\,pc.)

Based on the photometric distance of $\approx 65\,$pc, the anticipated annual parallactic shift of $\approx 15\,{\rm mas}$ is well below the astrometric accuracy of our data. Moreover the UKIRT exposure and the NIRI exposures were taken around the same time of year providing a six year baseline while further minimizing the parallactic shift. The proper motion of the source between these two exposures with respect to the field stars is $-120\pm30\,{\rm mas/yr}$, and $200\pm30\,{\rm mas/yr}$ along the RA and DEC axes respectively. Further details of our NIR astrometry are given in Appendix B5. Combined with the measured proper motion and its uncertainties, the corresponding tangential velocity is $73\pm14\,{\rm km}\,{\rm s}^{-1}$. This makes \jjj\ likely member of the thin disk population, based on the kinematic criteria of \citet{dupuy2012}.

\section{Discussion}
\subsection{Emission mechanism}
Brown dwarf radio emission falls into two phenomenological categories: (a) impulsive highly polarized emission from the cyclotron maser instability \citep[ECMI; ][]{hallinan2007,hallinan2008,kao2018,route2016a,route2016b}, and (b) quasi-quiescent emission with low polarization fraction that is attributed to incoherent gyrosynchrotron emission \citep{berger2001,williams2015,osten2006}.
Adopting the photometric distance of $d=65\,{\rm pc}$, the brightness temperature of the emitter in \jjj\, is $T_{\rm b} \approx 10^{15}\,{\rm K}x_\ast^{-2}$, where $x_\ast$ is the radius of the emitter in units of the characteristic brown dwarf radius of $7\times 10^9\,{\rm cm}$.  The high brightness temperature and circular fraction summarily rules out all incoherent emission mechanism. We therefore interpret the observed radio emission as ECMI. 

\subsection{Radio energetics \& temporal variation}
Circularly polarized radio emission in BDs is driven by magnetospheric acceleration processes \citep{hallinan2008,hallinan2015}, whose luminosity need not be rigidly related to the NIR luminosity that is determined by surface temperature and atmospheric composition. As such, the first radio selected BD in a flux-limited survey is likely to be more radio-luminous than the NIR-selected population.

Adopting the photometric distance of $d_{\rm pc} = 65$, the time-averaged (8-hr exposure) radio spectral luminosity in our 2018 detection is $\approx 5\times 10^{15}\,{\rm erg\,s}^{-1}\,{\rm Hz}^{-1}$. 
For comparison, highly polarized radio emission from previous T-dwarfs have only been detected to have time-averaged spectral luminosities below $\sim 10^{13}\,{\rm ergs}\,{\rm s}^{-1}\,{\rm Hz}^{-1}$ \citep{kao2018,kao2020,williams2013}. However, the brightest short-duration pulses from T-dwarfs typically last tens of seconds and attain a spectral luminosity of $\sim 10^{15}\,{\rm ergs}\,{\rm s}^{-1}\,{\rm Hz}^{-1}$ \citep{route2016a,route2016b}. Such values are comparable to the $8-$hr averaged value for \jjj.

To search for short radio bursts and any signature of rotation modulation, we extracted the radio light curve of \jjj\, from our 2018 detection at varying temporal cadences (Fig. \ref{fig:lc}). The light curves do not show a clear sign of periodicity and we see no evidence of intense short-duration bursts that could account for a significant fraction of the quasi-quiescent radio luminosity. We also computed a windowed FFT of the curves, as well as a Lomb-Scargle periodogram (Fig. \ref{fig:ps}). Again, we did not detect an unambiguous signature of periodicity (more details are available in the Appendix). 

\subsection{Special geometry or unusually luminous?}
That large distance-scale to \jjj\, is unusual given that it is the first radio-selected BD in a flux-limited survey. We explore two scenarios that may give \jjj\, its unusually large time-averaged spectral luminosity.
The scenarios also correspond to the two acceleration mechanisms that are known to operate in the Jovian magnetosphere and postulated to operate in BD magnetospheres--- breakdown of co-rotation between the plasma and the magnetic field \citep{nichols2012,turnpenney2017}, and a sub-Alfv\'{e}nic interaction with an orbiting companion \citep{turnpenney2018, saur2013}.

{\em Co-rotation breakdown: } One possibility is that there is nothing particular about \jjj\, when compared to other radio-loud T-dwarfs and that its high time-averaged spectral luminosity is merely a result of a special viewing geometry.\footnote{In that case, we recommend the qualified `R' be dropped from the name.} Based on the reference case of solar system planetary radio emissions, auroral radio emission is expected to primarily occur at high magnetic latitudes \citep{zarka1998}. It is also expected to be beamed along the surface of a cone aligned with its axis parallel to the ambient magnetic field and a large opening angle \citep{zarka1998,melrose1982,treumann}. In the hypothetical case of perfect axial symmetry, there will not be any rotation modulation of the observed emission. Real magnetospheres possess some azimuthal anomaly and/or a misalignment between the rotation and magnetic axes \citep{russell2010}. In the general case, the anomaly leads to a strong rotational modulation of the observed emission. The resulting emission typically appears pulsed and the pulse pattern repeats at the rotation period \citep[see for e.g.][]{hallinan2007,route2016b,kao2016,kao2018}. However, in specific geometries (equator-on view for instance), the emission may be visible over most rotational phases.\footnote{See \citet{hess2011,pineda2017} for examples of radio signatures of rotational modulation.}
Our discovery technique is biased towards finding systems with such a geometry because we blindly search for brown dwarfs in 8-hour exposure images. This scenario may explain our non-detection of periodicity due to the absence of a pulsed rotational modulation. In addition, the radio non-detection in one of the 2018 exposures could be the result of intrinsic variability expected in masers. For comparison, Jovian ECMI emission shows variability between epoch of factors of several \citep{zarka2004}. 

{\em Sub-Alfv\'{e}nic interaction:} Alternatively, the electrodynamic engine in \jjj\, maybe particularly luminous as it is driven by interaction with a close-by and/or large companion. In this case, the radiation is only beamed towards the Earth during specific combinations of the orbital phases of the companion and the rotational phase of the primary, similar to the visibility of the Io-related Jovian emission. This beaming geometry could account for the non-detection in the 2018-12-14 LOFAR exposure. 

The occurrence rate of planets around cold brown-dwarfs is currently not well constrained \citep{he2017}. Nevertheless, a rough constraint on the companion's size may be obtained by scaling up the Jupiter-Io benchmark to meet the observed radio luminosity.
Taking the emission bandwidth of \jjj\, to be $200\,$MHz for an estimate, the lower limit on the isotropic luminosity in the radio band is $10^{24}\,{\rm erg\,s}^{-1}$. The Jupiter-Io system generates an average radio power of $\sim 10^{17}\,{\rm erg}\,{\rm s}^{-1}$ \citep{zarka2004, zarka1998}. Assuming that the radio emission from \jjj\, is beamed into a solid angle of $0.16\,{\rm sr}$ as is the case for Io-driven Jovian emission \citep{zarka2004}, the observed emission is $10^{5}$ times more luminous than the Jupiter-Io system. Assuming the same interaction Mach-number as Io's interaction and the same geometric factors, the Poynting flux from the interaction scales as $R^2_{\rm obs}vB^2$ \citep{zarka2007,saur2013,turnpenney2018}, where  $R$ is the effective radius of the companion, $v$ is the relative velocity between the co-rotating magnetic field and the orbiting companion, and $B$ is the magnetic field at the radius of the companion. If we adopt a rotation period for \jjj\, of 2\,hours that is comparable to other radio-loud T-dwarfs \citep{kao2018}, and a surface field strength of $0.1\,{\rm kG}$ and the same orbital distance as that of Io, the necessary power can be achieved if the companion presents an obstruction of radius $\approx 0.25\,R_{\rm J}$. Because a dipolar field evolves with distance as $d^{-3}$, the term $vB^2$ in the expression for radio power evolves steeply as $d^{-5}$. Hence, this scenario admits a wide range of companion sizes. 
\subsection{Outlook}
The total power in the auroral current on \jjj\, can be further constrained by optical recombination-line observations. Taking the radio power to be $1\%$ of the kinetic power in the auroral electrons \cite{zarka2007, lamy2011} yields a total kinetic power of $10^{26}\,{\rm ergs}\,{\rm s}^{-1}$ which should be detectable in Balmer line emission for instance. In addition, a parallax measurement is necessary to secure a distance estimate. This is especially true since any close companion will contaminate the NIR flux of the object and produce erroneous photometric distance estimates.

The two scenarios presented above can be tested with radio data. 
If the companion driven emission is the true scenario, then continued radio monitoring should reveal signatures of periodicity at the orbital period of the companion. Such observation will, however, prove challenging due to the large inherent variation in maser luminosity.
If on the other hand, the special geometric alignment scenario is correct, then a search for BDs in short exposure radio images made with existing LOFAR data must reveal a large underlying population of bursts from BDs that do not have a special geometric alignment with respect to the Earth.

We end by noting that the \jjj\, is not only the first radio selected BD, but the low frequency of observation means that the magnetic field at the emitter is comparable to that anticipated in gas-giant exoplanets \citep{yadav2017,cauley2019,reiners2010}. Our discovery therefore bodes well for radio detections of exoplanet magnetospheres.

\acknowledgments
JRC thanks the Nederlandse Organisatie voor Wetenschappelijk Onderzoek (NWO) for support via the Talent Programme Veni grant. The authors thank Prof. Gregg Hallinan for commenting on the manuscript. This paper is based on data obtained with the International LOFAR Telescope (obs. IDs 691360 and 658492 as part of the LoTSS survey and obs. ID 763257 awarded to proposal LC13\textunderscore 021). LOFAR is the Low Frequency Array designed and constructed by ASTRON. It has observing, data processing, and data storage facilities in several countries, that are owned by various parties (each with their own funding sources), and that are collectively operated by the ILT foundation under a joint scientific policy. The ILT resources have benefitted from the following recent major funding sources: CNRS-INSU, Observatoire de Paris and Universit\'{e} d'Orl\'{e}ans, France; BMBF, MIWF-NRW, MPG, Germany; Science Foundation Ireland (SFI), Department of Business, Enterprise and Innovation (DBEI), Ireland; NWO, The Netherlands; The Science and Technology Facilities Council, UK. This research made use of the Dutch national e-infrastructure with support of the SURF Cooperative (e-infra 180169) and the LOFAR e-infra group. The J\"{u}lich LOFAR Long Term Archive and the German LOFAR network are both coordinated and operated by the J\"{u}lich Supercomputing Centre (JSC), and computing resources on the supercomputer JUWELS at JSC were provided by the Gauss Centre for Supercomputing e.V. (grant CHTB00) through the John von Neumann Institute for Computing (NIC). This research made use of the University of Hertfordshire high-performance computing facility and the LOFAR-UK computing facility located at the University of Hertfordshire and supported by STFC [ST/P000096/1], and of the Italian LOFAR IT computing infrastructure supported and operated by INAF, and by the Physics Department of Turin university (under an agreement with Consorzio Interuniversitario per la Fisica Spaziale) at the C3S Supercomputing Centre, Italy. The paper is based on observations obtained at the international Gemini Observatory (DDT proposal DT-2019B-014), a program of NSF's NOIRLab, which is managed by the Association of Universities for Research in Astronomy (AURA) under a cooperative agreement with the National Science Foundation on behalf of the Gemini Observatory partnership: the National Science Foundation (United States), National Research Council (Canada), Agencia Nacional de Investigaci\'{o}n y Desarrollo (Chile), Ministerio de Ciencia, Tecnología e Innovación (Argentina), Minist\'{e}rio da Ci\^{e}ncia, Tecnologia, Inova\c{c}\~{o}es e Comunica\c{c}\~{o}es (Brazil), and Korea Astronomy and Space Science Institute (Republic of Korea). {\em Software: } \texttt{python3}, \texttt{numpy}, \texttt{scipy}, \texttt{astropy}, \texttt{matplotlib}. {\em Facilities: } LOFAR, Gemini-North, WISE, UKIRT, Hale telescope.

%
\begin{figure*}
    \centering
    \includegraphics[width=\linewidth]{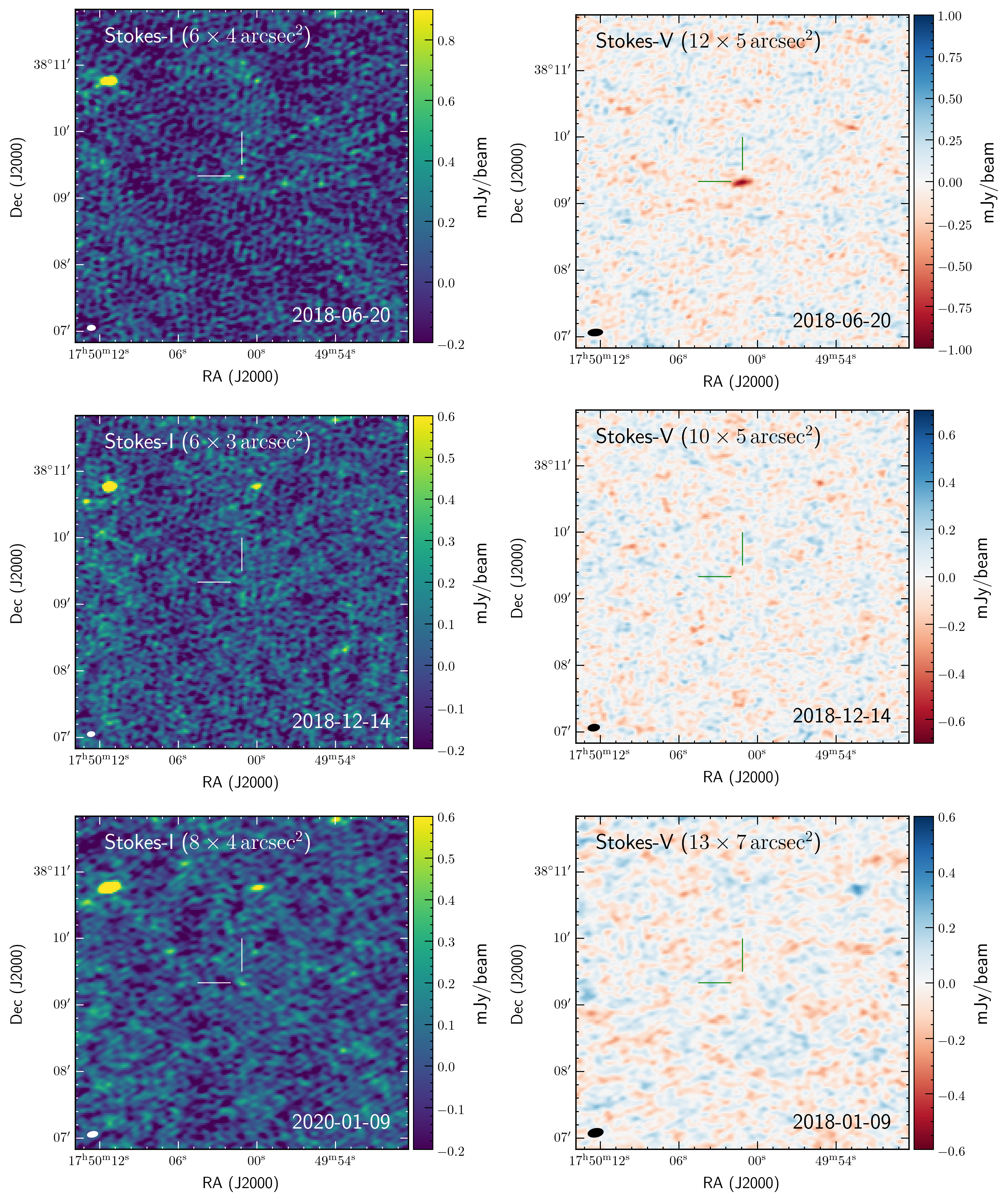}
    \caption{Radio detections and non-detections of \jjj\ with LOFAR. Left column shows Stokes-I (total intensity) images and right column shows Stokes-V (circularly polarized intensity) images made with Brigg's weighting with a factor of $-0.5$ and $0$ respectively. The observation dates and beam sizes are annotated. The position of \jjj\, is marked with cross-hairs that are $30\arcsec$ long. The images are $5\arcmin$ in size.}
    \label{fig:radio_montage}
\end{figure*}
\begin{figure*}
\centering
\includegraphics[width=\linewidth]{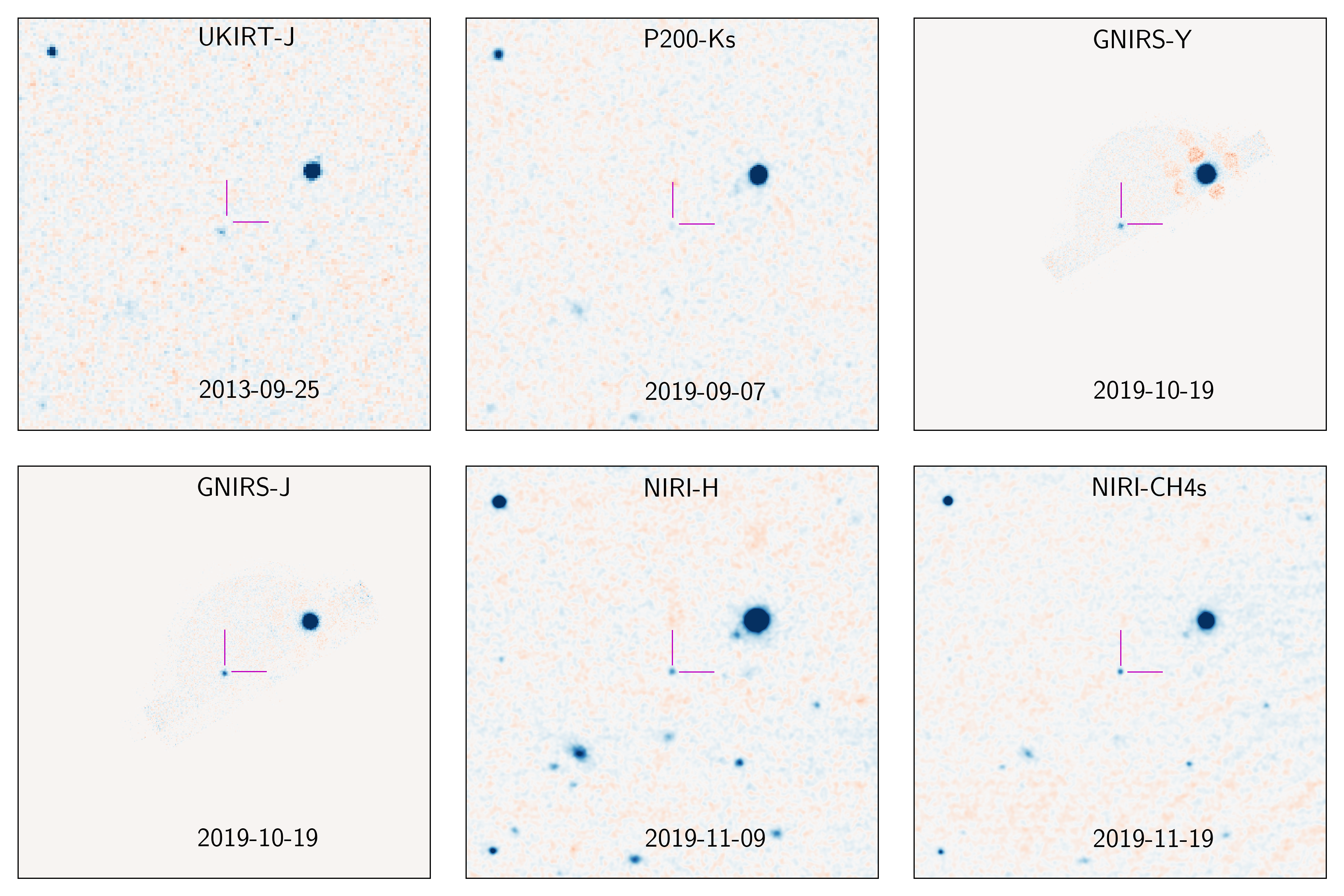}
\caption{NIR images of the field around \jjj. Because the images have disparate plate-scales, they have all been been convolved with a Gaussian kernel with FWHM of $0\arcsec.5$. Instrument names, filters and observation dates are in the annotated text. The images are $1\arcmin$ in size. The cross hairs are 5 arc-seconds long and point to the position of \jjj\, in the NIRI-CH4s exposure. Colour-scale runs from $-25\times$MAD (red) to $+25\times$MAD(blue), where MAD is the median absolute deviation and zero is denoted by white. Extracted source magnitudes and positions are given in Table \ref{tab:phot-astro}. The roughly semicircular field of view of the GNIRS keyhole is smaller than the image dimension. The red circular patches in the GNIRS-Yband image are at the 1\% level and are artifacts of flat-fielding.}
\label{fig:nir_montage}
\end{figure*}
\begin{table*}[]
    \centering
    \begin{tabular}{|c|c|c|c|c|}
        \hline
        Obs. date    & Telescope / Instrument        & Band      & Flux density  & Position \\ \hline
        2013-09-25  & UKIRT             & $J$-MK     & $19^{\rm m}.2(4)$ & 17:50:01.18(2), +38:09:18.5(2) \\
        2018-06-20  & LOFAR / HBA            & 144\,MHz  & 1.1(2) / -1.0(2)\,mJy      &  17:50:01.15(5), 38:09:19.6(8) \\
        2018-12-14  & LOFAR / HBA            & 144\,MHz  & 0.1(1) / -0.08(7)\,mJy  &   Non-detection       \\
        2020-01-09  & LOFAR / HBA            & 144\,MHz  &     0.4(1) / -0.05(7)\,mJy  &     Marginal detection ($4\sigma$)      \\
        
        2019-09-07  & Hale / WIRC   & $Ks$        &$19^{\rm m}.2(4)$ &   Marginal detection ($3\sigma$)        \\
        2019-10-19  & Gemini-N / GNIRS  & $J$-MK     &$19^{\rm m}.1(1)$& 17:50:01.13(2), +38:09:19.5(2)
          \\
        2019-10-19  & Gemini-N / GNIRS  & $Y$-MK     &$20^{\rm m}.4(1)$ & 17:50:01.12(2), +38:09:19.4(2)          \\
        2019-11-09  & Gemini-N / NIRI   & $H$-MK     &$19^{\rm m}.9(1)$ & 17:50:01.13(2), +38:09:19.5(2)          \\
        2019-11-13  & Genini-N / NIRI   & \ch     &$19^{\rm m}.3(1)$&  17:50:01.12(2), +38:09:19.6(2)         \\
        Various (co-add) & un-WISE & W1 & $18^{\rm m}.8(2)$ & 17:50:01(1), +38:09:19.7(7) \\
        Various (co-add) & un-WISE & W2 & $17^{\rm m}.2(2)$ &17:50:01(1), +38:09:20.3(7) \\ \hline
    \end{tabular}
    \caption{Photometry and astrometry of \jjj\, in J2000. Radio flux densities are given for Stokes-I and Stokes-V emission. Magnitudes are in the Vega system. Numbers in parenthesis give the error on the last significant digit. Errors in magnitude only include formal errors from aperture photometry and do not include systematic photometric errors (see appendix for further details).}
    \label{tab:phot-astro}
\end{table*}
\begin{figure}
    \centering
    \includegraphics[width=0.7\linewidth]{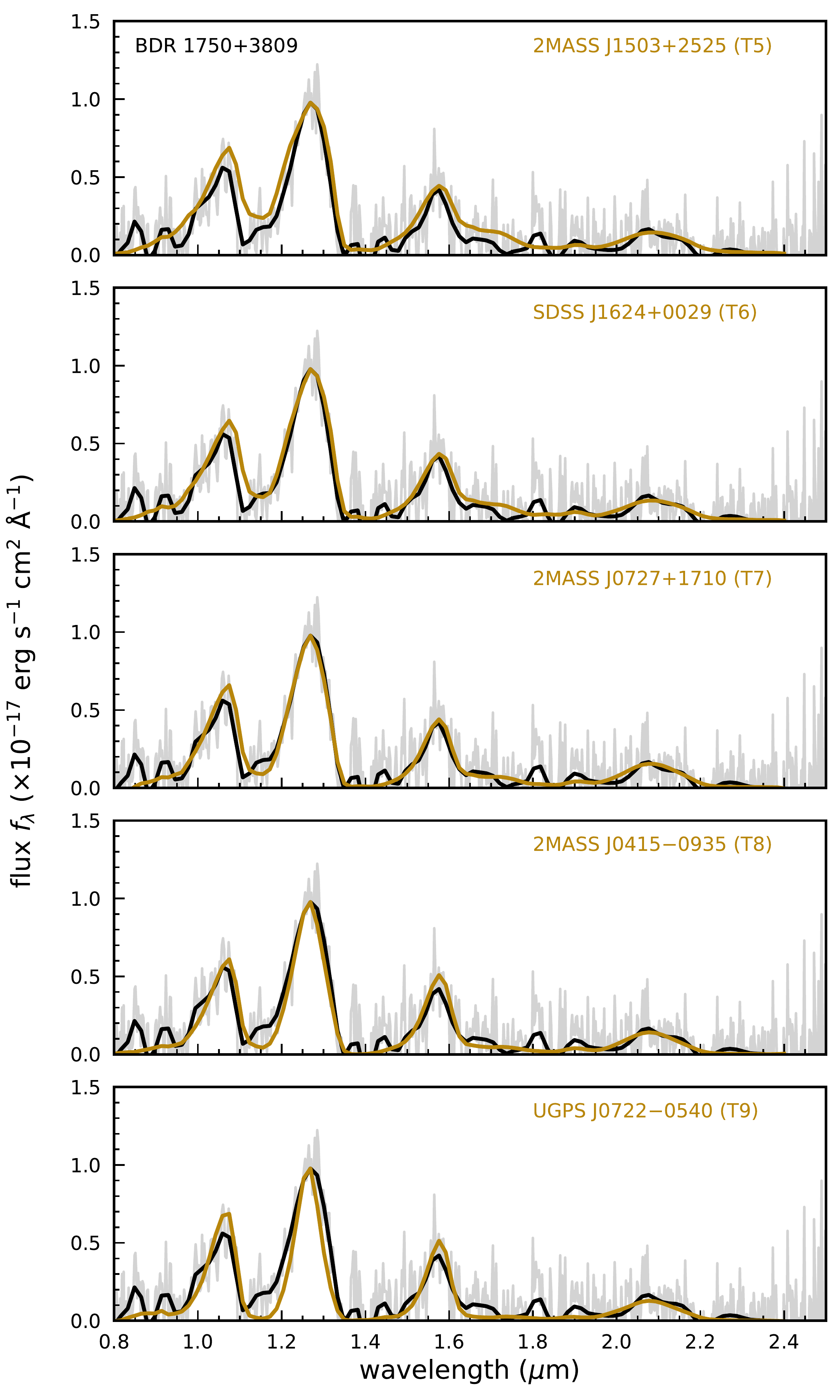}
    \caption{\normalsize Near-IR spectrum of \jjj\ (black)
    compared to T~dwarf spectral standards (tan) from
    \citet{2006ApJ...637.1067B} and \citet{2011ApJ...743...50C}. All these spectra are smoothed to $R\sim50$ for comparison and the reduced spectrum of \jjj\ prior to the smoothing is shown as grey. We have flux-calibrated the spectrum of \jjj\ based on its J-band magnitude from UKIRT (Table~\ref{tab:phot-astro}) and normalized spectral standards by their peak fluxes in J band.\label{fig:plot-spectra}}
    \end{figure}

\begin{figure*}
    \centering
    \includegraphics[width=\linewidth]{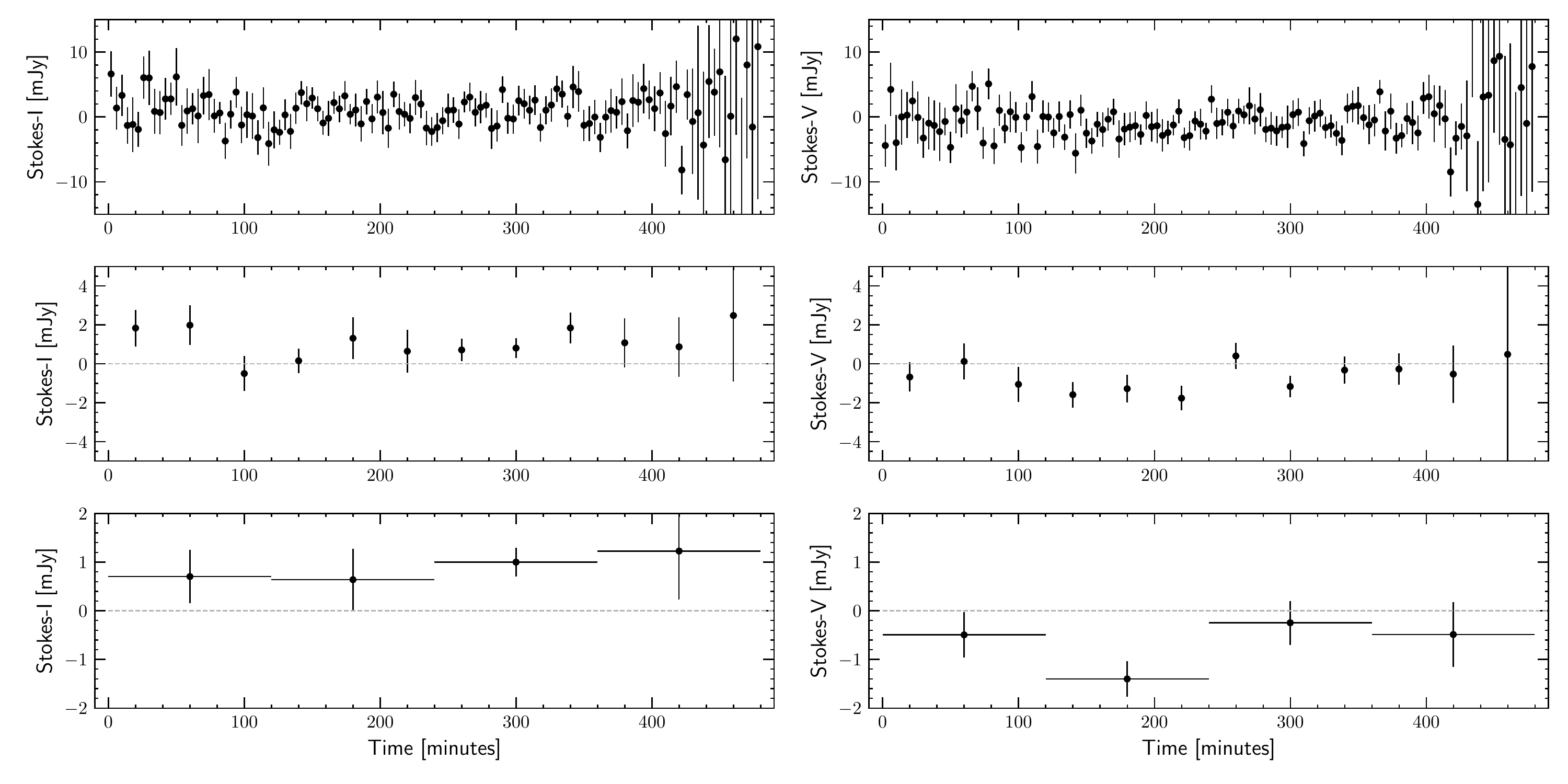}
    \caption{Stokes-I and Stokes-V light curves of \jjj\,(left and right hand columns) from the 2018 detection (see Fig. \ref{fig:radio_montage}) at varying temporal resolutions of $4^{\rm m}$, $40^{\rm m}$, and $120^{\rm m}$ (top to bottom). The light curves show that the emission has a stable brightness, with no obvious bright bursts.}
    \label{fig:lc}
\end{figure*}
\bibliography{sample63}{}

\begin{thebibliography}{}
\expandafter\ifx\csname natexlab\endcsname\relax\def\natexlab#1{#1}\fi
\providecommand{\url}[1]{\href{#1}{#1}}
\providecommand{\dodoi}[1]{doi:~\href{http://doi.org/#1}{\nolinkurl{#1}}}
\providecommand{\doeprint}[1]{\href{http://ascl.net/#1}{\nolinkurl{http://ascl.net/#1}}}
\providecommand{\doarXiv}[1]{\href{https://arxiv.org/abs/#1}{\nolinkurl{https://arxiv.org/abs/#1}}}

\bibitem[{{Bastian} {et~al.}(2000){Bastian}, {Dulk}, \&
  {Leblanc}}]{bastian2000}
{Bastian}, T.~S., {Dulk}, G.~A., \& {Leblanc}, Y. 2000, \apj, 545, 1058,
  \dodoi{10.1086/317864}

\bibitem[{{Berger} {et~al.}(2001){Berger}, {Ball}, {Becker}, {Clarke}, {Frail},
  {Fukuda}, {Hoffman}, {Mellon}, {Momjian}, {Murphy}, {Teng}, {Woodruff},
  {Zauderer}, \& {Zavala}}]{berger2001}
{Berger}, E., {Ball}, S., {Becker}, K.~M., {et~al.} 2001, \nat, 410, 338.
\newblock \doarXiv{astro-ph/0102301}

\bibitem[{Burgasser {et~al.}(2006)Burgasser, {Geballe}, {Leggett},
  {Kirkpatrick}, \& {Golimowski}}]{2006ApJ...637.1067B}
Burgasser, A.~J., {Geballe}, T.~R., {Leggett}, S.~K., {Kirkpatrick}, J.~D., \&
  {Golimowski}, D.~A. 2006, \apj, 637, 1067, \dodoi{10.1086/498563}

\bibitem[{{Burningham} {et~al.}(2016){Burningham}, {Hardcastle}, {Nichols},
  {Casewell}, {Littlefair}, {Stark}, {Burleigh}, {Metchev}, {Tannock}, {van
  Weeren}, {Williams}, \& {Wynn}}]{burningham2016}
{Burningham}, B., {Hardcastle}, M., {Nichols}, J.~D., {et~al.} 2016, \mnras,
  463, 2202, \dodoi{10.1093/mnras/stw2065}

\bibitem[{{Callingham} {et~al.}(2019){Callingham}, {Vedantham}, {Pope},
  {Shimwell}, \& {the LoTSS Team}}]{2019RNAAS...3...37C}
{Callingham}, J.~R., {Vedantham}, H.~K., {Pope}, B.~J.~S., {Shimwell}, T.~W.,
  \& {the LoTSS Team}. 2019, Research Notes of the American Astronomical
  Society, 3, 37, \dodoi{10.3847/2515-5172/ab07c3}

\bibitem[{{Callingham} {et~al.}(2020){Callingham}, {Vedantham}, {Shimwell}, \&
  {Pope}}]{joe-pop}
{Callingham}, J.~R., {Vedantham}, H.~K., {Shimwell}, T.~W., \& {Pope}, B.~J.~S.
  2020, Nature astronomy \,(under review), xxx, xxx

\bibitem[{{Cauley} {et~al.}(2019){Cauley}, {Shkolnik}, {Llama}, \&
  {Lanza}}]{cauley2019}
{Cauley}, P.~W., {Shkolnik}, E.~L., {Llama}, J., \& {Lanza}, A.~F. 2019, Nature
  Astronomy, 3, 1128, \dodoi{10.1038/s41550-019-0840-x}

\bibitem[{{Chambers} {et~al.}(2016){Chambers}, {Magnier}, {Metcalfe},
  {Flewelling}, {Huber}, {Waters}, {Denneau}, {Draper}, {Farrow}, {Finkbeiner},
  {Holmberg}, {Koppenhoefer}, {Price}, {Rest}, {Saglia}, {Schlafly}, {Smartt},
  {Sweeney}, {Wainscoat}, {Burgett}, {Chastel}, {Grav}, {Heasley}, {Hodapp},
  {Jedicke}, {Kaiser}, {Kudritzki}, {Luppino}, {Lupton}, {Monet}, {Morgan},
  {Onaka}, {Shiao}, {Stubbs}, {Tonry}, {White}, {Ba{\~n}ados}, {Bell},
  {Bender}, {Bernard}, {Boegner}, {Boffi}, {Botticella}, {Calamida},
  {Casertano}, {Chen}, {Chen}, {Cole}, {Deacon}, {Frenk}, {Fitzsimmons},
  {Gezari}, {Gibbs}, {Goessl}, {Goggia}, {Gourgue}, {Goldman}, {Grant},
  {Grebel}, {Hambly}, {Hasinger}, {Heavens}, {Heckman}, {Henderson}, {Henning},
  {Holman}, {Hopp}, {Ip}, {Isani}, {Jackson}, {Keyes}, {Koekemoer}, {Kotak},
  {Le}, {Liska}, {Long}, {Lucey}, {Liu}, {Martin}, {Masci}, {McLean}, {Mindel},
  {Misra}, {Morganson}, {Murphy}, {Obaika}, {Narayan}, {Nieto-Santisteban},
  {Norberg}, {Peacock}, {Pier}, {Postman}, {Primak}, {Rae}, {Rai}, {Riess},
  {Riffeser}, {Rix}, {R{\"o}ser}, {Russel}, {Rutz}, {Schilbach}, {Schultz},
  {Scolnic}, {Strolger}, {Szalay}, {Seitz}, {Small}, {Smith}, {Soderblom},
  {Taylor}, {Thomson}, {Taylor}, {Thakar}, {Thiel}, {Thilker}, {Unger},
  {Urata}, {Valenti}, {Wagner}, {Walder}, {Walter}, {Watters}, {Werner},
  {Wood-Vasey}, \& {Wyse}}]{ps1}
{Chambers}, K.~C., {Magnier}, E.~A., {Metcalfe}, N., {et~al.} 2016, arXiv
  e-prints, arXiv:1612.05560.
\newblock \doarXiv{1612.05560}

\bibitem[{{Christensen} {et~al.}(2009){Christensen}, {Holzwarth}, \&
  {Reiners}}]{christensen}
{Christensen}, U.~R., {Holzwarth}, V., \& {Reiners}, A. 2009, \nat, 457, 167,
  \dodoi{10.1038/nature07626}

\bibitem[{{Cushing} {et~al.}(2004){Cushing}, {Vacca}, \&
  {Rayner}}]{2004PASP..116..362C}
{Cushing}, M.~C., {Vacca}, W.~D., \& {Rayner}, J.~T. 2004, \pasp, 116, 362,
  \dodoi{10.1086/382907}

\bibitem[{Cushing {et~al.}(2011)Cushing, {Kirkpatrick}, {Gelino}, {Griffith},
  {Skrutskie}, {Mainzer}, {Marsh}, {Beichman}, {Burgasser}, {Prato}, {Simcoe},
  {Marley}, {Saumon}, {Freedman}, {Eisenhardt}, \&
  {Wright}}]{2011ApJ...743...50C}
Cushing, M.~C., {Kirkpatrick}, J.~D., {Gelino}, C.~R., {et~al.} 2011, \apj,
  743, 50, \dodoi{10.1088/0004-637X/743/1/50}

\bibitem[{{Cutri} \& {et al.}(2013)}]{wise}
{Cutri}, R.~M., \& {et al.} 2013, VizieR Online Data Catalog, II/328

\bibitem[{{De} {et~al.}(2020){De}, {Hankins}, {Kasliwal}, {Moore}, {Ofek},
  {Adams}, {Ashley}, {Babul}, {Bagdasaryan}, {Burdge}, {Burnham}, {Dekany},
  {Declacroix}, {Galla}, {Greffe}, {Hale}, {Jencson}, {Lau}, {Mahabal},
  {McKenna}, {Sharma}, {Shopbell}, {Smith}, {Soon}, {Sokoloski}, {Soria}, \&
  {Travouillon}}]{De2020}
{De}, K., {Hankins}, M.~J., {Kasliwal}, M.~M., {et~al.} 2020, \pasp, 132,
  025001, \dodoi{10.1088/1538-3873/ab6069}

\bibitem[{{Deeg} \& {Belmonte}(2018)}]{exoplanet-handbook}
{Deeg}, H.~J., \& {Belmonte}, J.~A. 2018, {Handbook of Exoplanets},
  \dodoi{10.1007/978-3-319-55333-7}

\bibitem[{{Dupuy} \& {Liu}(2012)}]{dupuy2012}
{Dupuy}, T.~J., \& {Liu}, M.~C. 2012, \apjs, 201, 19,
  \dodoi{10.1088/0067-0049/201/2/19}

\bibitem[{{Dye} {et~al.}(2018){Dye}, {Lawrence}, {Read}, {Fan}, {Kerr},
  {Varricatt}, {Furnell}, {Edge}, {Irwin}, {Hambly}, {Lucas}, {Almaini},
  {Chambers}, {Green}, {Hewett}, {Liu}, {McGreer}, {Best}, {Zhang}, {Sutorius},
  {Froebrich}, {Magnier}, {Hasinger}, {Lederer}, {Bold}, \& {Tedds}}]{uhsdr1}
{Dye}, S., {Lawrence}, A., {Read}, M.~A., {et~al.} 2018, \mnras, 473, 5113,
  \dodoi{10.1093/mnras/stx2622}

\bibitem[{{Elias} {et~al.}(2006){Elias}, {Joyce}, {Liang}, {Muller}, {Hileman},
  \& {George}}]{gnirs}
{Elias}, J.~H., {Joyce}, R.~R., {Liang}, M., {et~al.} 2006, in Society of
  Photo-Optical Instrumentation Engineers (SPIE) Conference Series, Vol. 6269,
  Society of Photo-Optical Instrumentation Engineers (SPIE) Conference Series,
  62694C, \dodoi{10.1117/12.671817}

\bibitem[{{Fegley} \& {Lodders}(1996)}]{fegley1996}
{Fegley}, Bruce, J., \& {Lodders}, K. 1996, \apjl, 472, L37,
  \dodoi{10.1086/310356}

\bibitem[{{Gaia Collaboration} {et~al.}(2018){Gaia Collaboration}, {Brown},
  {Vallenari}, {Prusti}, {de Bruijne}, {Babusiaux}, {Bailer-Jones}, {Biermann},
  {Evans}, {Eyer}, {Jansen}, {Jordi}, {Klioner}, {Lammers}, {Lindegren},
  {Luri}, {Mignard}, {Panem}, {Pourbaix}, {Randich}, {Sartoretti}, {Siddiqui},
  {Soubiran}, {van Leeuwen}, {Walton}, {Arenou}, {Bastian}, {Cropper},
  {Drimmel}, {Katz}, {Lattanzi}, {Bakker}, {Cacciari}, {Casta{\~n}eda},
  {Chaoul}, {Cheek}, {De Angeli}, {Fabricius}, {Guerra}, {Holl}, {Masana},
  {Messineo}, {Mowlavi}, {Nienartowicz}, {Panuzzo}, {Portell}, {Riello},
  {Seabroke}, {Tanga}, {Th{\'e}venin}, {Gracia-Abril}, {Comoretto},
  {Garcia-Reinaldos}, {Teyssier}, {Altmann}, {Andrae}, {Audard},
  {Bellas-Velidis}, {Benson}, {Berthier}, {Blomme}, {Burgess}, {Busso},
  {Carry}, {Cellino}, {Clementini}, {Clotet}, {Creevey}, {Davidson}, {De
  Ridder}, {Delchambre}, {Dell'Oro}, {Ducourant},
  {Fern{\'a}ndez-Hern{\'a}ndez}, {Fouesneau}, {Fr{\'e}mat}, {Galluccio},
  {Garc{\'\i}a-Torres}, {Gonz{\'a}lez-N{\'u}{\~n}ez}, {Gonz{\'a}lez-Vidal},
  {Gosset}, {Guy}, {Halbwachs}, {Hambly}, {Harrison}, {Hern{\'a}ndez},
  {Hestroffer}, {Hodgkin}, {Hutton}, {Jasniewicz}, {Jean-Antoine-Piccolo},
  {Jordan}, {Korn}, {Krone-Martins}, {Lanzafame}, {Lebzelter}, {L{\"o}ffler},
  {Manteiga}, {Marrese}, {Mart{\'\i}n-Fleitas}, {Moitinho}, {Mora}, {Muinonen},
  {Osinde}, {Pancino}, {Pauwels}, {Petit}, {Recio-Blanco}, {Richards},
  {Rimoldini}, {Robin}, {Sarro}, {Siopis}, {Smith}, {Sozzetti}, {S{\"u}veges},
  {Torra}, {van Reeven}, {Abbas}, {Abreu Aramburu}, {Accart}, {Aerts},
  {Altavilla}, {{\'A}lvarez}, {Alvarez}, {Alves}, {Anderson}, {Andrei},
  {Anglada Varela}, {Antiche}, {Antoja}, {Arcay}, {Astraatmadja}, {Bach},
  {Baker}, {Balaguer-N{\'u}{\~n}ez}, {Balm}, {Barache}, {Barata}, {Barbato},
  {Barblan}, {Barklem}, {Barrado}, {Barros}, {Barstow}, {Bartholom{\'e}
  Mu{\~n}oz}, {Bassilana}, {Becciani}, {Bellazzini}, {Berihuete}, {Bertone},
  {Bianchi}, {Bienaym{\'e}}, {Blanco-Cuaresma}, {Boch}, {Boeche}, {Bombrun},
  {Borrachero}, {Bossini}, {Bouquillon}, {Bourda}, {Bragaglia}, {Bramante},
  {Breddels}, {Bressan}, {Brouillet}, {Br{\"u}semeister}, {Brugaletta},
  {Bucciarelli}, {Burlacu}, {Busonero}, {Butkevich}, {Buzzi}, {Caffau},
  {Cancelliere}, {Cannizzaro}, {Cantat-Gaudin}, {Carballo}, {Carlucci},
  {Carrasco}, {Casamiquela}, {Castellani}, {Castro-Ginard}, {Charlot},
  {Chemin}, {Chiavassa}, {Cocozza}, {Costigan}, {Cowell}, {Crifo}, {Crosta},
  {Crowley}, {Cuypers}, {Dafonte}, {Damerdji}, {Dapergolas}, {David}, {David},
  {de Laverny}, {De Luise}, {De March}, {de Martino}, {de Souza}, {de Torres},
  {Debosscher}, {del Pozo}, {Delbo}, {Delgado}, {Delgado}, {Di Matteo},
  {Diakite}, {Diener}, {Distefano}, {Dolding}, {Drazinos}, {Dur{\'a}n},
  {Edvardsson}, {Enke}, {Eriksson}, {Esquej}, {Eynard Bontemps}, {Fabre},
  {Fabrizio}, {Faigler}, {Falc{\~a}o}, {Farr{\`a}s Casas}, {Federici},
  {Fedorets}, {Fernique}, {Figueras}, {Filippi}, {Findeisen}, {Fonti},
  {Fraile}, {Fraser}, {Fr{\'e}zouls}, {Gai}, {Galleti}, {Garabato},
  {Garc{\'\i}a-Sedano}, {Garofalo}, {Garralda}, {Gavel}, {Gavras}, {Gerssen},
  {Geyer}, {Giacobbe}, {Gilmore}, {Girona}, {Giuffrida}, {Glass}, {Gomes},
  {Granvik}, {Gueguen}, {Guerrier}, {Guiraud}, {Guti{\'e}rrez-S{\'a}nchez},
  {Haigron}, {Hatzidimitriou}, {Hauser}, {Haywood}, {Heiter}, {Helmi}, {Heu},
  {Hilger}, {Hobbs}, {Hofmann}, {Holland}, {Huckle}, {Hypki}, {Icardi},
  {Jan{\ss}en}, {Jevardat de Fombelle}, {Jonker}, {Juh{\'a}sz}, {Julbe},
  {Karampelas}, {Kewley}, {Klar}, {Kochoska}, {Kohley}, {Kolenberg},
  {Kontizas}, {Kontizas}, {Koposov}, {Kordopatis}, {Kostrzewa-Rutkowska},
  {Koubsky}, {Lambert}, {Lanza}, {Lasne}, {Lavigne}, {Le Fustec}, {Le
  Poncin-Lafitte}, {Lebreton}, {Leccia}, {Leclerc}, {Lecoeur-Taibi},
  {Lenhardt}, {Leroux}, {Liao}, {Licata}, {Lindstr{\o}m}, {Lister}, {Livanou},
  {Lobel}, {L{\'o}pez}, {Managau}, {Mann}, {Mantelet}, {Marchal}, {Marchant},
  {Marconi}, {Marinoni}, {Marschalk{\'o}}, {Marshall}, {Martino}, {Marton},
  {Mary}, {Massari}, {Matijevi{\v{c}}}, {Mazeh}, {McMillan}, {Messina},
  {Michalik}, {Millar}, {Molina}, {Molinaro}, {Moln{\'a}r}, {Montegriffo},
  {Mor}, {Morbidelli}, {Morel}, {Morris}, {Mulone}, {Muraveva}, {Musella},
  {Nelemans}, {Nicastro}, {Noval}, {O'Mullane}, {Ord{\'e}novic},
  {Ord{\'o}{\~n}ez-Blanco}, {Osborne}, {Pagani}, {Pagano}, {Pailler},
  {Palacin}, {Palaversa}, {Panahi}, {Pawlak}, {Piersimoni}, {Pineau}, {Plachy},
  {Plum}, {Poggio}, {Poujoulet}, {Pr{\v{s}}a}, {Pulone}, {Racero}, {Ragaini},
  {Rambaux}, {Ramos-Lerate}, {Regibo}, {Reyl{\'e}}, {Riclet}, {Ripepi}, {Riva},
  {Rivard}, {Rixon}, {Roegiers}, {Roelens}, {Romero-G{\'o}mez}, {Rowell},
  {Royer}, {Ruiz-Dern}, {Sadowski}, {Sagrist{\`a} Sell{\'e}s}, {Sahlmann},
  {Salgado}, {Salguero}, {Sanna}, {Santana-Ros}, {Sarasso}, {Savietto},
  {Schultheis}, {Sciacca}, {Segol}, {Segovia}, {S{\'e}gransan}, {Shih},
  {Siltala}, {Silva}, {Smart}, {Smith}, {Solano}, {Solitro}, {Sordo}, {Soria
  Nieto}, {Souchay}, {Spagna}, {Spoto}, {Stampa}, {Steele},
  {Steidelm{\"u}ller}, {Stephenson}, {Stoev}, {Suess}, {Surdej}, {Szabados},
  {Szegedi-Elek}, {Tapiador}, {Taris}, {Tauran}, {Taylor}, {Teixeira},
  {Terrett}, {Teyssand ier}, {Thuillot}, {Titarenko}, {Torra Clotet}, {Turon},
  {Ulla}, {Utrilla}, {Uzzi}, {Vaillant}, {Valentini}, {Valette}, {van Elteren},
  {Van Hemelryck}, {van Leeuwen}, {Vaschetto}, {Vecchiato}, {Veljanoski},
  {Viala}, {Vicente}, {Vogt}, {von Essen}, {Voss}, {Votruba}, {Voutsinas},
  {Walmsley}, {Weiler}, {Wertz}, {Wevers}, {Wyrzykowski}, {Yoldas},
  {{\v{Z}}erjal}, {Ziaeepour}, {Zorec}, {Zschocke}, {Zucker}, {Zurbach}, \&
  {Zwitter}}]{gaia}
{Gaia Collaboration}, {Brown}, A.~G.~A., {Vallenari}, A., {et~al.} 2018, \aap,
  616, A1, \dodoi{10.1051/0004-6361/201833051}

\bibitem[{{Hallinan} {et~al.}(2008){Hallinan}, {Antonova}, {Doyle}, {Bourke},
  {Lane}, \& {Golden}}]{hallinan2008}
{Hallinan}, G., {Antonova}, A., {Doyle}, J.~G., {et~al.} 2008, \apj, 684, 644,
  \dodoi{10.1086/590360}

\bibitem[{{Hallinan} {et~al.}(2013){Hallinan}, {Sirothia}, {Antonova},
  {Ishwara-Chand ra}, {Bourke}, {Doyle}, {Hartman}, \& {Golden}}]{hallinan2013}
{Hallinan}, G., {Sirothia}, S.~K., {Antonova}, A., {et~al.} 2013, \apj, 762,
  34, \dodoi{10.1088/0004-637X/762/1/34}

\bibitem[{{Hallinan} {et~al.}(2007){Hallinan}, {Bourke}, {Lane}, {Antonova},
  {Zavala}, {Brisken}, {Boyle}, {Vrba}, {Doyle}, \& {Golden}}]{hallinan2007}
{Hallinan}, G., {Bourke}, S., {Lane}, C., {et~al.} 2007, \apjl, 663, L25,
  \dodoi{10.1086/519790}

\bibitem[{{Hallinan} {et~al.}(2015){Hallinan}, {Littlefair}, {Cotter},
  {Bourke}, {Harding}, {Pineda}, {Butler}, {Golden}, {Basri}, {Doyle}, {Kao},
  {Berdyugina}, {Kuznetsov}, {Rupen}, \& {Antonova}}]{hallinan2015}
{Hallinan}, G., {Littlefair}, S.~P., {Cotter}, G., {et~al.} 2015, \nat, 523,
  568, \dodoi{10.1038/nature14619}

\bibitem[{{Hancock} {et~al.}(2012){Hancock}, {Murphy}, {Gaensler}, {Hopkins},
  \& {Curran}}]{2012MNRAS.422.1812H}
{Hancock}, P.~J., {Murphy}, T., {Gaensler}, B.~M., {Hopkins}, A., \& {Curran},
  J.~R. 2012, \mnras, 422, 1812, \dodoi{10.1111/j.1365-2966.2012.20768.x}

\bibitem[{{Hancock} {et~al.}(2018){Hancock}, {Trott}, \&
  {Hurley-Walker}}]{2018PASA...35...11H}
{Hancock}, P.~J., {Trott}, C.~M., \& {Hurley-Walker}, N. 2018, \pasa, 35, e011,
  \dodoi{10.1017/pasa.2018.3}

\bibitem[{{He} {et~al.}(2017){He}, {Triaud}, \& {Gillon}}]{he2017}
{He}, M.~Y., {Triaud}, A. H.~M.~J., \& {Gillon}, M. 2017, \mnras, 464, 2687,
  \dodoi{10.1093/mnras/stw2391}

\bibitem[{{Hess} \& {Zarka}(2011)}]{hess2011}
{Hess}, S.~L.~G., \& {Zarka}, P. 2011, \aap, 531, A29,
  \dodoi{10.1051/0004-6361/201116510}

\bibitem[{{Hodapp} {et~al.}(2003){Hodapp}, {Jensen}, {Irwin}, {Yamada},
  {Chung}, {Fletcher}, {Robertson}, {Hora}, {Simons}, {Mays}, {Nolan}, {Bec},
  {Merrill}, \& {Fowler}}]{niri}
{Hodapp}, K.~W., {Jensen}, J.~B., {Irwin}, E.~M., {et~al.} 2003, \pasp, 115,
  1388, \dodoi{10.1086/379669}

\bibitem[{{Kao} {et~al.}(2019){Kao}, {Hallinan}, \& {Pineda}}]{kao2020}
{Kao}, M.~M., {Hallinan}, G., \& {Pineda}, J.~S. 2019, \mnras, 487, 1994,
  \dodoi{10.1093/mnras/stz1372}

\bibitem[{{Kao} {et~al.}(2016){Kao}, {Hallinan}, {Pineda}, {Escala},
  {Burgasser}, {Bourke}, \& {Stevenson}}]{kao2016}
{Kao}, M.~M., {Hallinan}, G., {Pineda}, J.~S., {et~al.} 2016, \apj, 818, 24,
  \dodoi{10.3847/0004-637X/818/1/24}

\bibitem[{{Kao} {et~al.}(2018){Kao}, {Hallinan}, {Pineda}, {Stevenson}, \&
  {Burgasser}}]{kao2018}
{Kao}, M.~M., {Hallinan}, G., {Pineda}, J.~S., {Stevenson}, D., \& {Burgasser},
  A. 2018, \apjs, 237, 25, \dodoi{10.3847/1538-4365/aac2d5}

\bibitem[{{Kirkpatrick} {et~al.}(1999){Kirkpatrick}, {Reid}, {Liebert},
  {Cutri}, {Nelson}, {Beichman}, {Dahn}, {Monet}, {Gizis}, \&
  {Skrutskie}}]{kirkpatrick1999}
{Kirkpatrick}, J.~D., {Reid}, I.~N., {Liebert}, J., {et~al.} 1999, \apj, 519,
  802, \dodoi{10.1086/307414}

\bibitem[{{Lacy} {et~al.}(2020){Lacy}, {Baum}, {Chandler}, {Chatterjee},
  {Clarke}, {Deustua}, {English}, {Farnes}, {Gaensler}, {Gugliucci},
  {Hallinan}, {Kent}, {Kimball}, {Law}, {Lazio}, {Marvil}, {Mao}, {Medlin},
  {Mooley}, {Murphy}, {Myers}, {Osten}, {Richards}, {Rosolowsky}, {Rudnick},
  {Schinzel}, {Sivakoff}, {Sjouwerman}, {Taylor}, {White}, {Wrobel},
  {Andernach}, {Beasley}, {Berger}, {Bhatnager}, {Birkinshaw}, {Bower},
  {Brandt}, {Brown}, {Burke-Spolaor}, {Butler}, {Comerford}, {Demorest}, {Fu},
  {Giacintucci}, {Golap}, {G{\"u}th}, {Hales}, {Hiriart}, {Hodge}, {Horesh},
  {Ivezi{\'c}}, {Jarvis}, {Kamble}, {Kassim}, {Liu}, {Loinard}, {Lyons},
  {Masters}, {Mezcua}, {Moellenbrock}, {Mroczkowski}, {Nyland},
  {O{\textquoteright}Dea}, {O{\textquoteright}Sullivan}, {Peters}, {Radford},
  {Rao}, {Robnett}, {Salcido}, {Shen}, {Sobotka}, {Witz}, {Vaccari}, {van
  Weeren}, {Vargas}, {Williams}, \& {Yoon}}]{vlass}
{Lacy}, M., {Baum}, S.~A., {Chandler}, C.~J., {et~al.} 2020, \pasp, 132,
  035001, \dodoi{10.1088/1538-3873/ab63eb}

\bibitem[{{Lamy} {et~al.}(2011){Lamy}, {Cecconi}, {Zarka}, {Canu}, {Schippers},
  {Kurth}, {Mutel}, {Gurnett}, {Menietti}, \& {Louarn}}]{lamy2011}
{Lamy}, L., {Cecconi}, B., {Zarka}, P., {et~al.} 2011, Journal of Geophysical
  Research (Space Physics), 116, A04212, \dodoi{10.1029/2010JA016195}

\bibitem[{{Lazio} {et~al.}(2004){Lazio}, {Farrell}, {Dietrick}, {Greenlees},
  {Hogan}, {Jones}, \& {Hennig}}]{lazio2004}
{Lazio}, T.~Joseph, W., {Farrell}, W.~M., {Dietrick}, J., {et~al.} 2004, \apj,
  612, 511, \dodoi{10.1086/422449}

\bibitem[{{Leggett} {et~al.}(2017){Leggett}, {Tremblin}, {Esplin}, {Luhman}, \&
  {Morley}}]{leggett2017}
{Leggett}, S.~K., {Tremblin}, P., {Esplin}, T.~L., {Luhman}, K.~L., \&
  {Morley}, C.~V. 2017, \apj, 842, 118, \dodoi{10.3847/1538-4357/aa6fb5}

\bibitem[{{Lenc} {et~al.}(2018){Lenc}, {Murphy}, {Lynch}, {Kaplan}, \&
  {Zhang}}]{lenc2018}
{Lenc}, E., {Murphy}, T., {Lynch}, C.~R., {Kaplan}, D.~L., \& {Zhang}, S.~N.
  2018, \mnras, 478, 2835, \dodoi{10.1093/mnras/sty1304}

\bibitem[{{Liu} {et~al.}(2008){Liu}, {Dupuy}, \& {Ireland}}]{liu2008}
{Liu}, M.~C., {Dupuy}, T.~J., \& {Ireland}, M.~J. 2008, \apj, 689, 436,
  \dodoi{10.1086/591837}

\bibitem[{{Lynch} {et~al.}(2017{\natexlab{a}}){Lynch}, {Lenc}, {Kaplan},
  {Murphy}, \& {Anderson}}]{lynch}
{Lynch}, C.~R., {Lenc}, E., {Kaplan}, D.~L., {Murphy}, T., \& {Anderson}, G.~E.
  2017{\natexlab{a}}, \apjl, 836, L30, \dodoi{10.3847/2041-8213/aa5ffd}

\bibitem[{{Lynch} {et~al.}(2017{\natexlab{b}}){Lynch}, {Murphy}, {Kaplan},
  {Ireland}, \& {Bell}}]{lynch2017}
{Lynch}, C.~R., {Murphy}, T., {Kaplan}, D.~L., {Ireland}, M., \& {Bell}, M.~E.
  2017{\natexlab{b}}, \mnras, 467, 3447, \dodoi{10.1093/mnras/stx354}

\bibitem[{{Melrose} \& {Dulk}(1982)}]{melrose1982}
{Melrose}, D.~B., \& {Dulk}, G.~A. 1982, \apj, 259, 844, \dodoi{10.1086/160219}

\bibitem[{{Nakajima} {et~al.}(2004){Nakajima}, {Tsuji}, \&
  {Yanagisawa}}]{nakajima2004}
{Nakajima}, T., {Tsuji}, T., \& {Yanagisawa}, K. 2004, \apj, 607, 499,
  \dodoi{10.1086/383299}

\bibitem[{{Nichols} {et~al.}(2012){Nichols}, {Burleigh}, {Casewell}, {Cowley},
  {Wynn}, {Clarke}, \& {West}}]{nichols2012}
{Nichols}, J.~D., {Burleigh}, M.~R., {Casewell}, S.~L., {et~al.} 2012, \apj,
  760, 59, \dodoi{10.1088/0004-637X/760/1/59}

\bibitem[{{Osten} {et~al.}(2006){Osten}, {Hawley}, {Bastian}, \&
  {Reid}}]{osten2006}
{Osten}, R.~A., {Hawley}, S.~L., {Bastian}, T.~S., \& {Reid}, I.~N. 2006, \apj,
  637, 518, \dodoi{10.1086/498345}

\bibitem[{{Pineda} {et~al.}(2017){Pineda}, {Hallinan}, \& {Kao}}]{pineda2017}
{Pineda}, J.~S., {Hallinan}, G., \& {Kao}, M.~M. 2017, \apj, 846, 75,
  \dodoi{10.3847/1538-4357/aa8596}

\bibitem[{Rayner {et~al.}(2003)Rayner, {Toomey}, {Onaka}, {Denault},
  {Stahlberger}, {Vacca}, {Cushing}, \& {Wang}}]{2003PASP..115..362R}
Rayner, J.~T., {Toomey}, D.~W., {Onaka}, P.~M., {et~al.} 2003, \pasp, 115, 362

\bibitem[{{Reiners} \& {Christensen}(2010)}]{reiners2010}
{Reiners}, A., \& {Christensen}, U.~R. 2010, \aap, 522, A13,
  \dodoi{10.1051/0004-6361/201014251}

\bibitem[{{Rosenthal} {et~al.}(1996){Rosenthal}, {Gurwell}, \&
  {Ho}}]{rosenthal1996}
{Rosenthal}, E.~D., {Gurwell}, M.~A., \& {Ho}, P. T.~P. 1996, \nat, 384, 243,
  \dodoi{10.1038/384243a0}

\bibitem[{{Route} \& {Wolszczan}(2016{\natexlab{a}})}]{route2016a}
{Route}, M., \& {Wolszczan}, A. 2016{\natexlab{a}}, \apj, 830, 85,
  \dodoi{10.3847/0004-637X/830/2/85}

\bibitem[{{Route} \& {Wolszczan}(2016{\natexlab{b}})}]{route2016b}
---. 2016{\natexlab{b}}, \apjl, 821, L21, \dodoi{10.3847/2041-8205/821/2/L21}

\bibitem[{{Russell} \& {Dougherty}(2010)}]{russell2010}
{Russell}, C.~T., \& {Dougherty}, M.~K. 2010, \ssr, 152, 251,
  \dodoi{10.1007/s11214-009-9621-7}

\bibitem[{{Saur} {et~al.}(2013){Saur}, {Grambusch}, {Duling}, {Neubauer}, \&
  {Simon}}]{saur2013}
{Saur}, J., {Grambusch}, T., {Duling}, S., {Neubauer}, F.~M., \& {Simon}, S.
  2013, \aap, 552, A119, \dodoi{10.1051/0004-6361/201118179}

\bibitem[{{Schlafly} {et~al.}(2019){Schlafly}, {Meisner}, \& {Green}}]{unwise}
{Schlafly}, E.~F., {Meisner}, A.~M., \& {Green}, G.~M. 2019, \apjs, 240, 30,
  \dodoi{10.3847/1538-4365/aafbea}

\bibitem[{{Schrijver} \& {Zwaan}(2008)}]{schrijver2008}
{Schrijver}, C.~J., \& {Zwaan}, C. 2008, {Solar and Stellar Magnetic Activity}
  (Cambridge University Press)

\bibitem[{{Schwenn}(2006)}]{schwenn2006}
{Schwenn}, R. 2006, Living Reviews in Solar Physics, 3, 2,
  \dodoi{10.12942/lrsp-2006-2}

\bibitem[{{Shimwell} {et~al.}(2017){Shimwell}, {R{\"o}ttgering}, {Best},
  {Williams}, {Dijkema}, {de Gasperin}, {Hardcastle}, {Heald}, {Hoang},
  {Horneffer}, {Intema}, {Mahony}, {Mandal}, {Mechev}, {Morabito}, {Oonk},
  {Rafferty}, {Retana-Montenegro}, {Sabater}, {Tasse}, {van Weeren},
  {Br{\"u}ggen}, {Brunetti}, {Chy{\.z}y}, {Conway}, {Haverkorn}, {Jackson},
  {Jarvis}, {McKean}, {Miley}, {Morganti}, {White}, {Wise}, {van Bemmel},
  {Beck}, {Brienza}, {Bonafede}, {Calistro Rivera}, {Cassano}, {Clarke},
  {Cseh}, {Deller}, {Drabent}, {van Driel}, {Engels}, {Falcke}, {Ferrari},
  {Fr{\"o}hlich}, {Garrett}, {Harwood}, {Heesen}, {Hoeft}, {Horellou},
  {Israel}, {Kapi{\'n}ska}, {Kunert-Bajraszewska}, {McKay}, {Mohan},
  {Orr{\'u}}, {Pizzo}, {Prandoni}, {Schwarz}, {Shulevski}, {Sipior}, {Smith},
  {Sridhar}, {Steinmetz}, {Stroe}, {Varenius}, {van der Werf}, {Zensus}, \&
  {Zwart}}]{lotss1}
{Shimwell}, T.~W., {R{\"o}ttgering}, H.~J.~A., {Best}, P.~N., {et~al.} 2017,
  \aap, 598, A104, \dodoi{10.1051/0004-6361/201629313}

\bibitem[{{Shimwell} {et~al.}(2019){Shimwell}, {Tasse}, {Hardcastle}, {Mechev},
  {Williams}, {Best}, {R{\"o}ttgering}, {Callingham}, {Dijkema}, {de Gasperin},
  {Hoang}, {Hugo}, {Mirmont}, {Oonk}, {Prandoni}, {Rafferty}, {Sabater},
  {Smirnov}, {van Weeren}, {White}, {Atemkeng}, {Bester}, {Bonnassieux},
  {Br{\"u}ggen}, {Brunetti}, {Chy{\.z}y}, {Cochrane}, {Conway}, {Croston},
  {Danezi}, {Duncan}, {Haverkorn}, {Heald}, {Iacobelli}, {Intema}, {Jackson},
  {Jamrozy}, {Jarvis}, {Lakhoo}, {Mevius}, {Miley}, {Morabito}, {Morganti},
  {Nisbet}, {Orr{\'u}}, {Perkins}, {Pizzo}, {Schrijvers}, {Smith}, {Vermeulen},
  {Wise}, {Alegre}, {Bacon}, {van Bemmel}, {Beswick}, {Bonafede}, {Botteon},
  {Bourke}, {Brienza}, {Calistro Rivera}, {Cassano}, {Clarke}, {Conselice},
  {Dettmar}, {Drabent}, {Dumba}, {Emig}, {En{\ss}lin}, {Ferrari}, {Garrett},
  {G{\'e}nova-Santos}, {Goyal}, {G{\"u}rkan}, {Hale}, {Harwood}, {Heesen},
  {Hoeft}, {Horellou}, {Jackson}, {Kokotanekov}, {Kondapally},
  {Kunert-Bajraszewska}, {Mahatma}, {Mahony}, {Mandal}, {McKean}, {Merloni},
  {Mingo}, {Miskolczi}, {Mooney}, {Nikiel-Wroczy{\'n}ski}, {O'Sullivan},
  {Quinn}, {Reich}, {Roskowi{\'n}ski}, {Rowlinson}, {Savini}, {Saxena},
  {Schwarz}, {Shulevski}, {Sridhar}, {Stacey}, {Urquhart}, {van der Wiel},
  {Varenius}, {Webster}, \& {Wilber}}]{lotss2}
{Shimwell}, T.~W., {Tasse}, C., {Hardcastle}, M.~J., {et~al.} 2019, \aap, 622,
  A1, \dodoi{10.1051/0004-6361/201833559}

\bibitem[{{Skrutskie} {et~al.}(2006){Skrutskie}, {Cutri}, {Stiening},
  {Weinberg}, {Schneider}, {Carpenter}, {Beichman}, {Capps}, {Chester},
  {Elias}, {Huchra}, {Liebert}, {Lonsdale}, {Monet}, {Price}, {Seitzer},
  {Jarrett}, {Kirkpatrick}, {Gizis}, {Howard}, {Evans}, {Fowler}, {Fullmer},
  {Hurt}, {Light}, {Kopan}, {Marsh}, {McCallon}, {Tam}, {Van Dyk}, \&
  {Wheelock}}]{2mass}
{Skrutskie}, M.~F., {Cutri}, R.~M., {Stiening}, R., {et~al.} 2006, \aj, 131,
  1163, \dodoi{10.1086/498708}

\bibitem[{{Stephens} \& {Leggett}(2004)}]{stephens2004}
{Stephens}, D.~C., \& {Leggett}, S.~K. 2004, \pasp, 116, 9,
  \dodoi{10.1086/381135}

\bibitem[{{Tinney} {et~al.}(2005){Tinney}, {Burgasser}, {Kirkpatrick}, \&
  {McElwain}}]{tinney2005}
{Tinney}, C.~G., {Burgasser}, A.~J., {Kirkpatrick}, J.~D., \& {McElwain}, M.~W.
  2005, \aj, 130, 2326, \dodoi{10.1086/491734}

\bibitem[{{Treumann}(2006)}]{treumann}
{Treumann}, R.~A. 2006, \aapr, 13, 229, \dodoi{10.1007/s00159-006-0001-y}

\bibitem[{{Turnpenney} {et~al.}(2018){Turnpenney}, {Nichols}, {Wynn}, \&
  {Burleigh}}]{turnpenney2018}
{Turnpenney}, S., {Nichols}, J.~D., {Wynn}, G.~A., \& {Burleigh}, M.~R. 2018,
  \apj, 854, 72, \dodoi{10.3847/1538-4357/aaa59c}

\bibitem[{{Turnpenney} {et~al.}(2017){Turnpenney}, {Nichols}, {Wynn}, \&
  {Casewell}}]{turnpenney2017}
{Turnpenney}, S., {Nichols}, J.~D., {Wynn}, G.~A., \& {Casewell}, S.~L. 2017,
  \mnras, 470, 4274, \dodoi{10.1093/mnras/stx1508}

\bibitem[{Vacca {et~al.}(2003)Vacca, {Cushing}, \&
  {Rayner}}]{2003PASP..115..389V}
Vacca, W.~D., {Cushing}, M.~C., \& {Rayner}, J.~T. 2003, \pasp, 115, 389

\bibitem[{{van Haarlem} {et~al.}(2013){van Haarlem}, {Wise}, {Gunst}, {Heald},
  {McKean}, {Hessels}, {de Bruyn}, {Nijboer}, {Swinbank}, {Fallows},
  {Brentjens}, {Nelles}, {Beck}, {Falcke}, {Fender}, {H{\"o}randel},
  {Koopmans}, {Mann}, {Miley}, {R{\"o}ttgering}, {Stappers}, {Wijers},
  {Zaroubi}, {van den Akker}, {Alexov}, {Anderson}, {Anderson}, {van Ardenne},
  {Arts}, {Asgekar}, {Avruch}, {Batejat}, {B{\"a}hren}, {Bell}, {Bell}, {van
  Bemmel}, {Bennema}, {Bentum}, {Bernardi}, {Best}, {B{\^\i}rzan}, {Bonafede},
  {Boonstra}, {Braun}, {Bregman}, {Breitling}, {van de Brink}, {Broderick},
  {Broekema}, {Brouw}, {Br{\"u}ggen}, {Butcher}, {van Cappellen}, {Ciardi},
  {Coenen}, {Conway}, {Coolen}, {Corstanje}, {Damstra}, {Davies}, {Deller},
  {Dettmar}, {van Diepen}, {Dijkstra}, {Donker}, {Doorduin}, {Dromer}, {Drost},
  {van Duin}, {Eisl{\"o}ffel}, {van Enst}, {Ferrari}, {Frieswijk}, {Gankema},
  {Garrett}, {de Gasperin}, {Gerbers}, {de Geus}, {Grie{\ss}meier}, {Grit},
  {Gruppen}, {Hamaker}, {Hassall}, {Hoeft}, {Holties}, {Horneffer}, {van der
  Horst}, {van Houwelingen}, {Huijgen}, {Iacobelli}, {Intema}, {Jackson},
  {Jelic}, {de Jong}, {Juette}, {Kant}, {Karastergiou}, {Koers}, {Kollen},
  {Kondratiev}, {Kooistra}, {Koopman}, {Koster}, {Kuniyoshi}, {Kramer},
  {Kuper}, {Lambropoulos}, {Law}, {van Leeuwen}, {Lemaitre}, {Loose}, {Maat},
  {Macario}, {Markoff}, {Masters}, {McFadden}, {McKay-Bukowski}, {Meijering},
  {Meulman}, {Mevius}, {Middelberg}, {Millenaar}, {Miller-Jones}, {Mohan},
  {Mol}, {Morawietz}, {Morganti}, {Mulcahy}, {Mulder}, {Munk}, {Nieuwenhuis},
  {van Nieuwpoort}, {Noordam}, {Norden}, {Noutsos}, {Offringa}, {Olofsson},
  {Omar}, {Orr{\'u}}, {Overeem}, {Paas}, {Pand ey-Pommier}, {Pandey}, {Pizzo},
  {Polatidis}, {Rafferty}, {Rawlings}, {Reich}, {de Reijer}, {Reitsma},
  {Renting}, {Riemers}, {Rol}, {Romein}, {Roosjen}, {Ruiter}, {Scaife}, {van
  der Schaaf}, {Scheers}, {Schellart}, {Schoenmakers}, {Schoonderbeek},
  {Serylak}, {Shulevski}, {Sluman}, {Smirnov}, {Sobey}, {Spreeuw}, {Steinmetz},
  {Sterks}, {Stiepel}, {Stuurwold}, {Tagger}, {Tang}, {Tasse}, {Thomas},
  {Thoudam}, {Toribio}, {van der Tol}, {Usov}, {van Veelen}, {van der Veen},
  {ter Veen}, {Verbiest}, {Vermeulen}, {Vermaas}, {Vocks}, {Vogt}, {de Vos},
  {van der Wal}, {van Weeren}, {Weggemans}, {Weltevrede}, {White}, {Wijnholds},
  {Wilhelmsson}, {Wucknitz}, {Yatawatta}, {Zarka}, {Zensus}, \& {van
  Zwieten}}]{lofar}
{van Haarlem}, M.~P., {Wise}, M.~W., {Gunst}, A.~W., {et~al.} 2013, \aap, 556,
  A2, \dodoi{10.1051/0004-6361/201220873}

\bibitem[{{Vedantham} {et~al.}(2020){Vedantham}, {Callingham}, {Shimwell},
  {Tasse}, {Pope}, {Bedell}, {Snellen}, {Best}, {Hardcastle}, {Haverkorn},
  {Mechev}, {O'Sullivan}, {R{\"o}ttgering}, \& {White}}]{gj1151}
{Vedantham}, H.~K., {Callingham}, J.~R., {Shimwell}, T.~W., {et~al.} 2020,
  Nature Astronomy, 4, 577, \dodoi{10.1038/s41550-020-1011-9}

\bibitem[{{Williams}(2018)}]{williams2018}
{Williams}, P. K.~G. 2018, {Radio Emission from Ultracool Dwarfs} ({Springer
  International Publishing}), 171, \dodoi{10.1007/978-3-319-55333-7_171}

\bibitem[{{Williams} {et~al.}(2013){Williams}, {Berger}, \&
  {Zauderer}}]{williams2013}
{Williams}, P. K.~G., {Berger}, E., \& {Zauderer}, B.~A. 2013, \apjl, 767, L30,
  \dodoi{10.1088/2041-8205/767/2/L30}

\bibitem[{{Williams} {et~al.}(2015){Williams}, {Casewell}, {Stark},
  {Littlefair}, {Helling}, \& {Berger}}]{williams2015}
{Williams}, P.~K.~G., {Casewell}, S.~L., {Stark}, C.~R., {et~al.} 2015, \apj,
  815, 64, \dodoi{10.1088/0004-637X/815/1/64}

\bibitem[{{Wilson} {et~al.}(2003){Wilson}, {Eikenberry}, {Henderson},
  {Hayward}, {Carson}, {Pirger}, {Barry}, {Brand l}, {Houck}, {Fitzgerald}, \&
  {Stolberg}}]{Wilson2003}
{Wilson}, J.~C., {Eikenberry}, S.~S., {Henderson}, C.~P., {et~al.} 2003, in
  Society of Photo-Optical Instrumentation Engineers (SPIE) Conference Series,
  Vol. 4841, Instrument Design and Performance for Optical/Infrared
  Ground-based Telescopes, ed. M.~{Iye} \& A.~F.~M. {Moorwood}, 451--458,
  \dodoi{10.1117/12.460336}

\bibitem[{{Wu} \& {Lee}(1979)}]{wu}
{Wu}, C.~S., \& {Lee}, L.~C. 1979, \apj, 230, 621, \dodoi{10.1086/157120}

\bibitem[{{Yadav} \& {Thorngren}(2017)}]{yadav2017}
{Yadav}, R.~K., \& {Thorngren}, D.~P. 2017, \apjl, 849, L12,
  \dodoi{10.3847/2041-8213/aa93fd}

\bibitem[{{Zarka}(1998)}]{zarka1998}
{Zarka}, P. 1998, \jgr, 103, 20159, \dodoi{10.1029/98JE01323}

\bibitem[{{Zarka}(2007)}]{zarka2007}
---. 2007, \planss, 55, 598, \dodoi{10.1016/j.pss.2006.05.045}

\bibitem[{{Zarka} {et~al.}(2004){Zarka}, {Cecconi}, \& {Kurth}}]{zarka2004}
{Zarka}, P., {Cecconi}, B., \& {Kurth}, W.~S. 2004, Journal of Geophysical
  Research (Space Physics), 109, A09S15, \dodoi{10.1029/2003JA010260}

\end{thebibliography}
\bibliographystyle{aasjournal}
\appendix
\section{Radio data}
\subsection{Data reduction}
We used the standard LoTSS pipeline for primary data reduction \citep{lotss1,lotss2}. An additional self-calibration step was applied in the direction of the target with a pipeline that is described in \citet{gj1151}.
All images were made with \texttt{wsclean} with Brigg's weighting. The images in Fig. \ref{fig:radio_montage} have a weighting factor of $-0.5$ for Stokes-I to suppress confusion from diffuse emission and sidelobe noise. The Stokes-V images do not suffer from these sources of confusion and have been made with a weighting factor of $0$ to maximize signal to noise ratio. The astrometric fits and flux density in Table \ref{tab:phot-astro} were determined from images made with a weighting factor of $-0.5$ to improve astrometric accuracy. 

We used the Background And Noise Estimator (\textsc{BANE}) and source finder \textsc{Aegean} \citep[v\,2.1.1;][]{2012MNRAS.422.1812H,2018PASA...35...11H} to measure the flux density and location of \jjj. Originally, we discovered \jjj~through a blind search for sources that were $>4\sigma$ in Stokes V emission, where $\sigma$ is the local rms noise \citep{2019RNAAS...3...37C,gj1151}. Once the position of the source was known, we applied the priorised fitting option of \textsc{Aegean} for the other epochs, which fits for both the PSF shape and flux density of \jjj. In the Stokes V images we searched for both positive and negative emission. 

The right source to the NE of the target has a peak Stokes-I flux density of $\sim 8.5\,$mJy and is undetected in the Stokes-V images with rms noise of 0.1\,mJy, suggesting that the Stokes I to V leakage is at the $\sim 1\%$ level of below in our images.

\subsection{Light curves}
Although the 8-hr exposure images have good $uv$ coverage, the short exposures suffer from sidelobe noise from in-field sources. We therefore modelled the visibilities of the in-field sources using the \texttt{update-model-column} option of \texttt{wsclean} and subtracted the model from the visibilities. To extract the light curves at the location of \jjj, the residual visibilities were then snapshot imaged at varying temporal cadences of with a Brigg's factor of $0$. 
\begin{figure}
    \centering
    \includegraphics[width=0.9\linewidth]{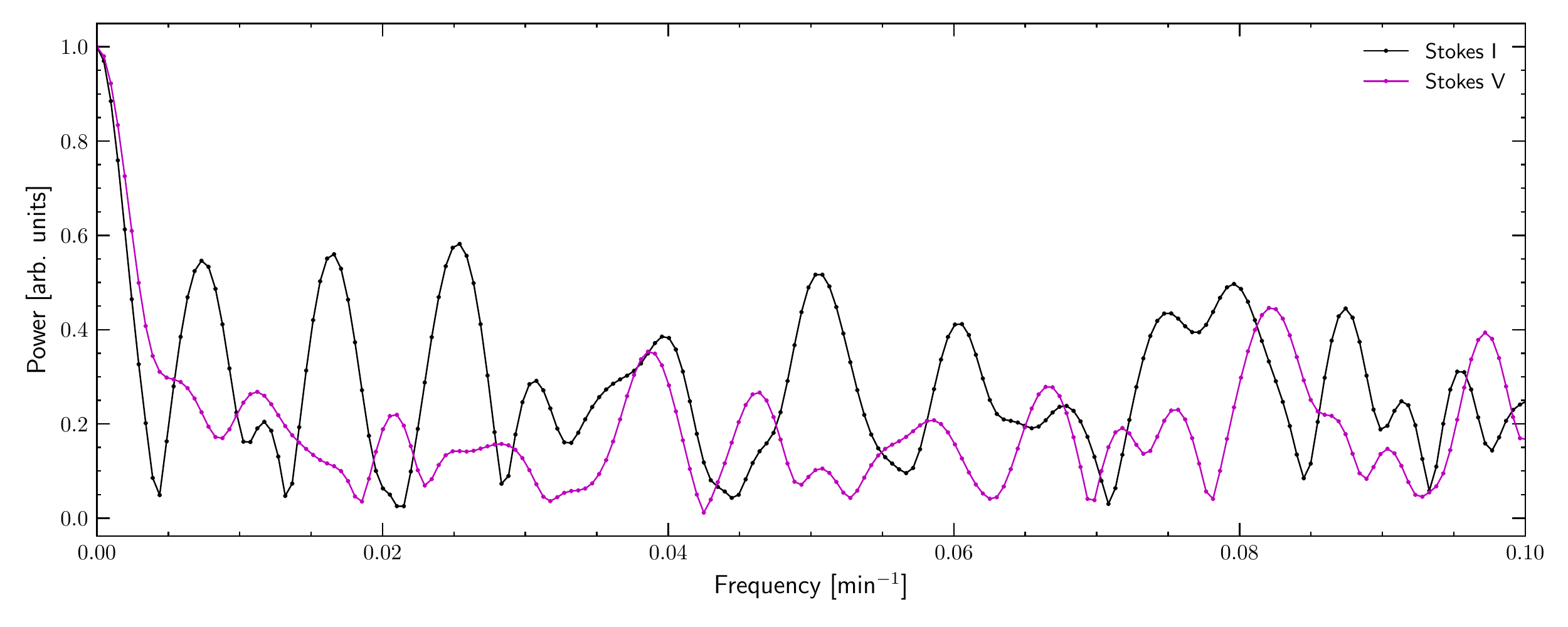}
    \includegraphics[width=0.9\linewidth]{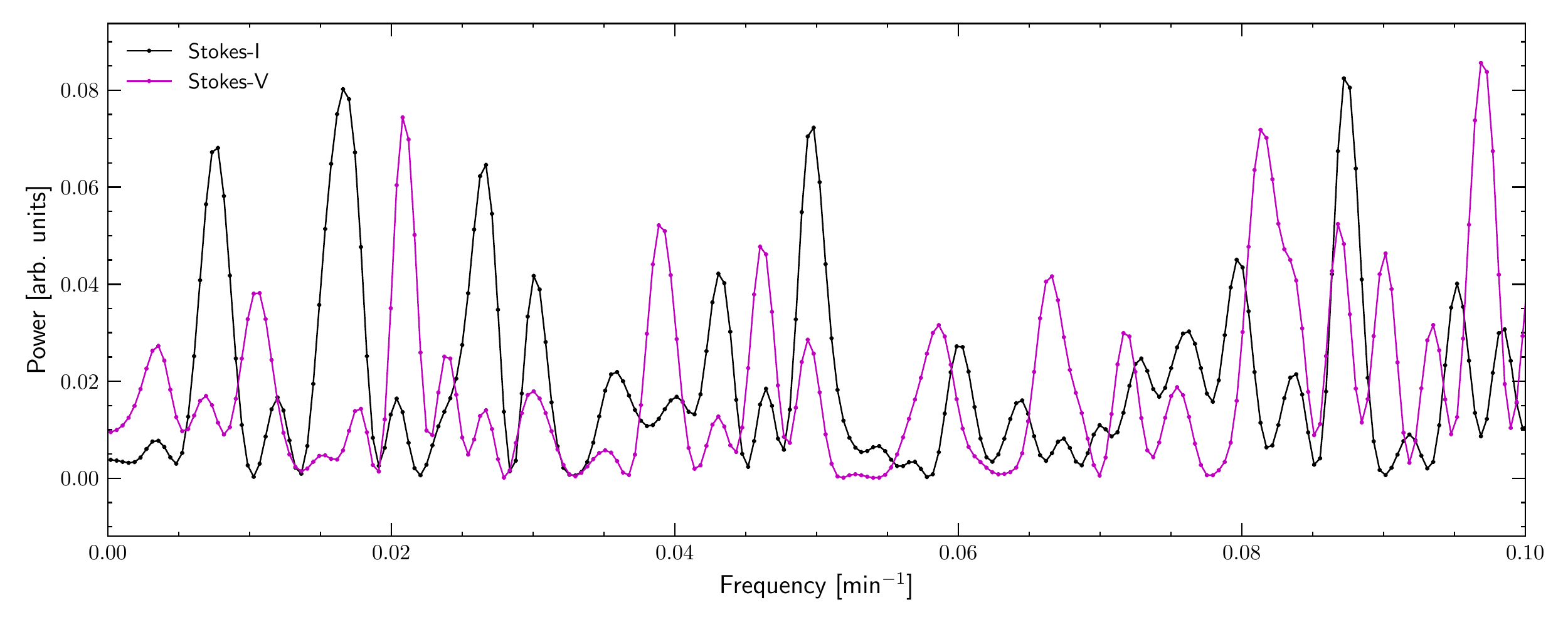}
    \caption{Power spectrum of temporal variations (top panel) and Lomb-Scargle periodogram (bottom panel). Light-curves at a $4\,$min cadence were used as input. A Hanning window was used to improve the point spread function of the FFT.} 
    \label{fig:ps}
\end{figure}
\subsection{Radio astrometry} The LoTSS astrometry is tied to the Pan-STARRS grid with an absolute astrometric error of $0\arcsec.2$ \citep{lotss1,lotss2}. In Table \ref{tab:phot-astro} we quote the error obtained by adding the formal error from our source finding in quadrature with the absolute astrometric error. Because \jjj\, is a faint source, its astrometric uncertainty in the radio is dominated by the formal error in finding the source centroid within the point spread function. For instance, the images used for astrometric determination from the 2018-06-20 exposure (Brigg's factor of $-0.5$) has a point spread function with major and minor axes of $4\arcsec \times 7\arcsec$. Adopting a geometric mean width of $5\arcsec .3$, the formal centroid-finding error for a $7\sigma$ source is $0\arcsec .76$. 
\section{NIR data reduction}
\subsection{GNIRS keyhole imaging}
There was light cloud cover and fog during the observations. Dark and bias current was subtracted from each exposure using custom python code applied to calibration images taken at the end of the night. The dome-flat frames were unusable due to improper illumination (cause unknown), hence we used the median combination of the dithered science exposures to make a sky-flat. The pixel centroid of Star\,A in each frame was determined using \texttt{sextractor}. The \texttt{FITS} header keywords \texttt{CRVAL1}, \texttt{CRPIX1}, \texttt{CRVAL2}, \texttt{CRPIX2} were modified to shift the frame so as to have Star\,A's position tied to its {\em Gaia} DR2 position \citep{gaia}. The plate scale and orientation could not be solved for with just one reference star, so we adopted the nominal values specified by the observatory. The resulting frames were resampled on to a common grid and median-combined using the \texttt{swarp} software.
\subsection{NIRI imaging}
As with the GNIRS exposures, the observing conditions did not allow for photometric calibration transfer from standard stars. The dark and bias currents were subtracted from each exposure using custom python code. The dome flats were found to be inadequate. So we used the dithered science exposures to construct a sky-flat which was applied in addition to the dome-flats. For each exposure, we then used \texttt{sextractor} to extract sources and \texttt{scamp} to solve for plate distortions up to third order while using the \texttt{USNO-B1} catalog as reference. Finally, we used \texttt{swarp} to re-sample the exposures on a common grid and median combine them.
\subsection{NIR photometry}
\begin{figure}
    \centering
    \includegraphics[width=0.6\linewidth]{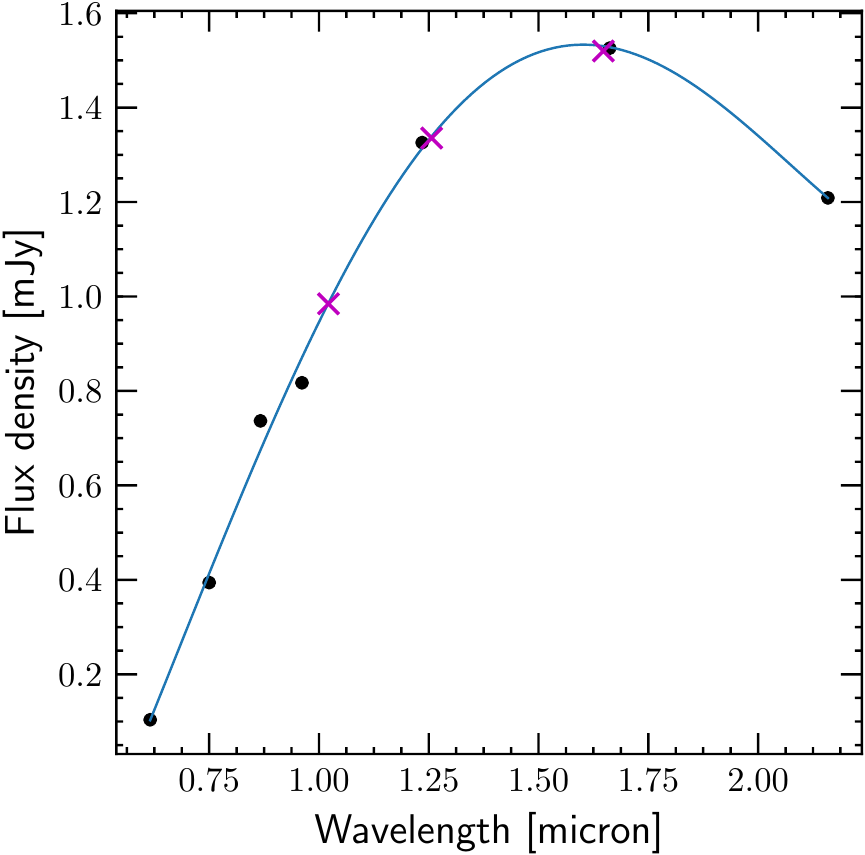}
    \caption{Measured (black dots) and estimated (magenta crosses) flux density of Star\,A (\sa). The estimates are based on a smooth polynomial fit (solid blue line) to the measurements.}
    \label{fig:star_A}
\end{figure}

Star A (2MASS\,J17500008+3809276) with a 2MASS $J$-band magnitude of $15^{\rm m}.22$ has measured flux densities in several optical, NIR and MIR bands (Fig. \ref{fig:star_A}). 
From a smooth polynomial fit to the star's flux densities (in Jansky units) measured by the Pan-STARRS and 2MASS surveys, we estimated the Vega magnitude of Star A in the MKO-systems $Y$, $J$, $H$ and \ch filters to be, respectively, $15^{\rm m}.78$, $15^{\rm m}.16$, $14^{\rm m}.58$ and $14^{\rm m}.61$. In doing so we assumed the zero-points of $2026\,{\rm Jy}$, $1545\,{\rm Jy}$, $1030\,{\rm Jy}$ and $1071\,{\rm Jy}$ respectively.
The fractional deviation of the Star A's photometric measurements in the near-infrared and the fit is within 1\%, which is smaller than the final photometric uncertainty (see below).
Based on its spectrum, the star is likely a mid M-dwarf (M3 or M4) which is not expected to have large spectral excursions in the NIR part of its spectrum. We checked individual exposures to make sure that the star did not display egregious flaring that would significantly affect its flux density in co-added images.
To determine the flux density (in ADC counts), we first computed the growth curve of 2MASS stars in the field my measuring their flux in varying apertures. The growth curves were averaged to yield the average growth curve. The small field of view of the GNIRS keyhole exposures meant that the only available 2MASS star in the field was Star\,A. We then measured the flux of the target in different apertures--- $1.0,\,1.25,\,1.5$ and so on until $2.5$ times the FWHM of point spread function. To determine the targets total flux, we fit the average growth curve to the target's growth curve measured with these apertures. We took the mean value of the fitted fluxes as the measured target flux and their dispersion as the formal flux density fitting error. The target and Star\,A's flux densities (in counts) were finally scaled to match Star\,A's measured flux with its model SED (Fig. \ref{fig:star_A}). 

The formal flux fitting errors were $0^{\rm m}.02-0^{\rm m}.05$ depending on the filter. We repeated the same photometric procedure on in-field 2MASS stars in our $H$-band and found our estimates to be differ from 2MASS estimates by about $0^{\rm m}.1$. We therefore conservatively adopted this value as the final error in our photometry.

Given the marginal detection in the $Ks$ band image the aperture flux with radii much larger than the seeing FWHM were severely affected by background estimation errors. We therefore measured the flux only in a single aperture whose radius was comparable to the seeing FWHM, instead of fitting the growth curves at various apertures. The photometry was referred to Star\,A as in the case of the Gemini observations. We note that a filter correction of $\approx -0^{\rm m}.2$ for a late T-dwarf \citep{stephens2004}, places the $K$ band magnitude of \jjj\, in the MKO system at $19^{\rm m}.4(4)$.  

\subsection{NIR spectroscopy}
We used the facility
near-IR spectrograph SpeX \citep{2003PASP..115..362R} in prism mode, obtaining 0.8--2.5~\micron\ spectra in a single order, with the 0.8\arcsec\ wide slit oriented at the parallactic angle. To acquire \jjj, we offset from Star A ($J=15.2$~mag) that lies 14\arcsec\ WNW (offsets of 12.25\arcsec\ east and 7.19\arcsec\ south from the 2MASS star). \jjj\ was nodded along the slit in an ABBA pattern, with individual exposure times of 180~sec, and observed over an airmass range of 1.3--2.0, resulting in a total on-source exposure time of 4320~sec. The telescope was guided
using the off-axis optical guide camera. We observed the A0~V star HD~165029 contemporaneously for flux and telluric calibration, interleaving observations of the science target and calibrator to minimize the airmass difference between the two. The spectra were reduced using version~4.1 of the SpeXtool software package \citep{2003PASP..115..389V,2004PASP..116..362C}.
\subsection{unWISE detection}
\begin{figure}
    \centering
    \includegraphics[width=0.7\linewidth]{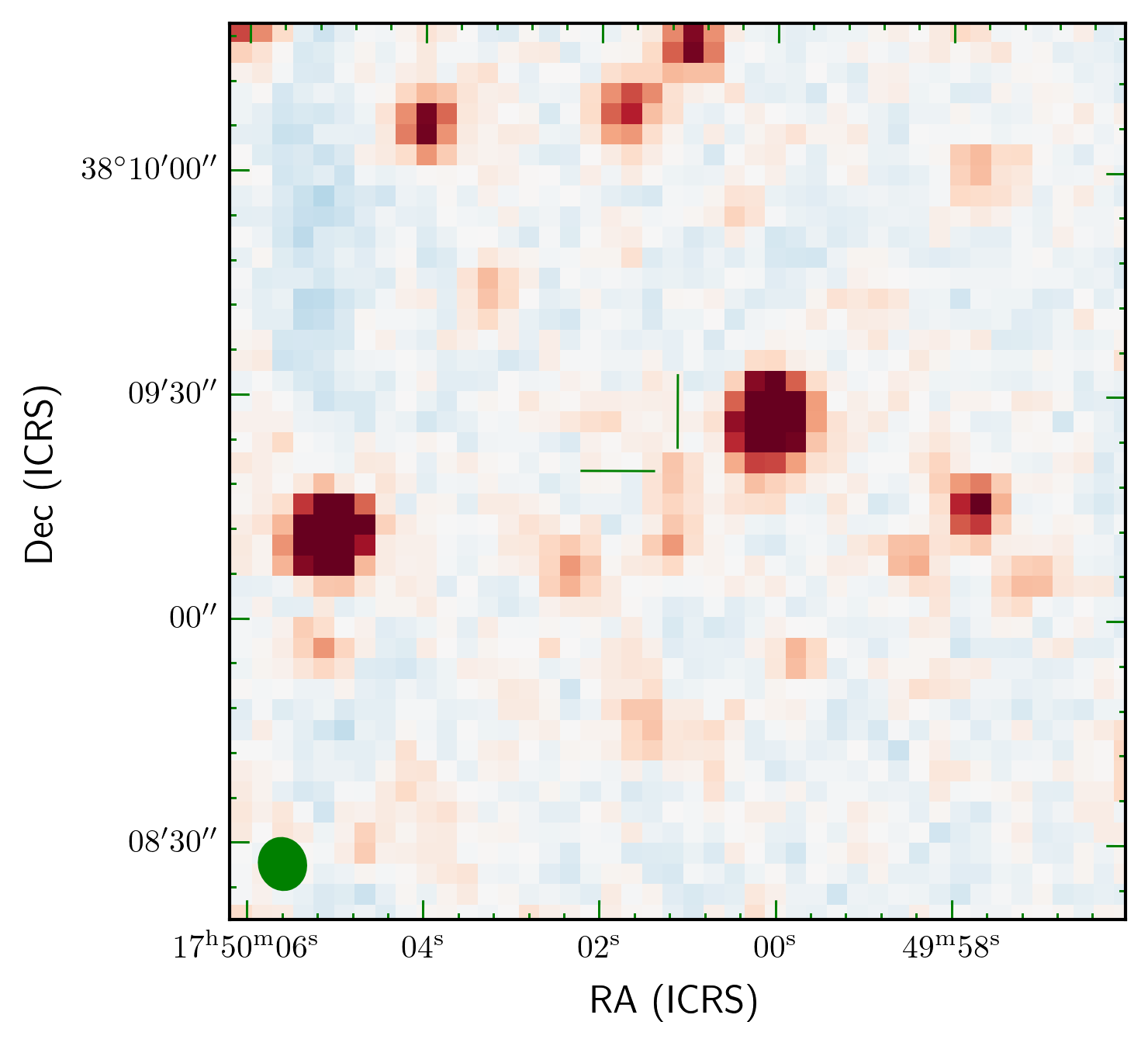}
    \caption{Image of the region around \jjj\, from the unWISE co-addition of WISE frames in filter W2 (centered on 4.5 microns). The plate-scale is $2\arcsec.75$ and the beam is shown as a green ellipse. The yellow cross-hairs are $10\arcsec$ in length and mark the position of \jjj\, from the NIRI-CH4s exposure (see table \ref{tab:phot-astro} and Fig. \ref{fig:nir_montage}). Colorscale runs from -15 to 15 median absolute deviation. }
    \label{fig:wise-w2}
\end{figure}

The original \texttt{AllWISE} catalogue does not have a source plausibly associated with \jjj\, even when accounting for proper motion. This catalogue had deliberately blurred point-spread function in the final co-added images. Recently \citet{unwise} have published `un-blurred' co-added images and extracted catalog. The W2-filter image is shown in Fig. \ref{fig:wise-w2} and we have reported the catalog flux in Table \ref{tab:phot-astro}. The catalog reports a detection of \jjj,\, at the $\approx 5\sigma$ level. It is flagged for the possible contamination from the wings of the bright source Star\,A to the North-West. Although there is almost of a decade that has elapsed within the exposures, the WISE point-spread function of $6\arcsec.7$ is large enough for the proper-motion to not affect the flux determination significantly.
\begin{figure}
    \centering
    \includegraphics[width=0.97\linewidth]{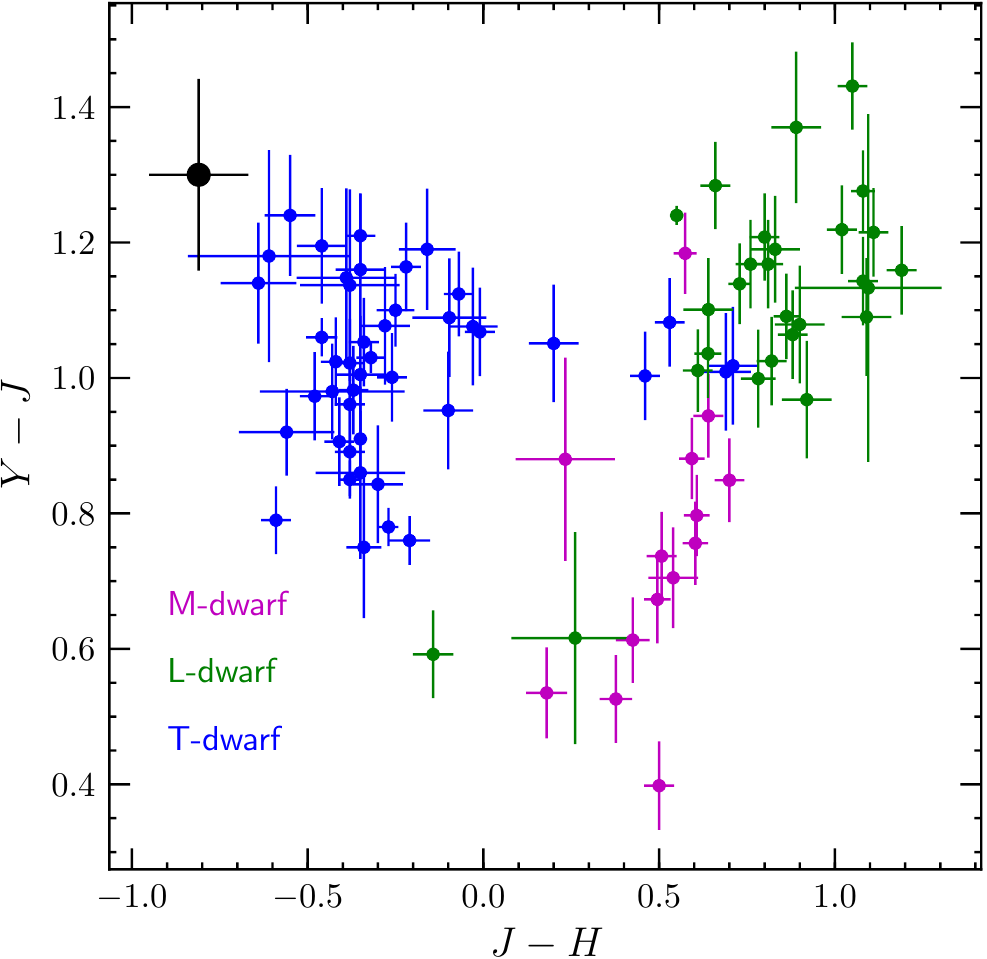}
    \caption{NIR colors of \jjj\,(shown in black) overplotted on the colours of M-, L- and T-dwarfs in the catalogue of \citet{dupuy2012}. The colors demonstrate that \jjj\,is a T-dwarf.}
    \label{fig:nir-color}
\end{figure}
\begin{figure}
    \centering
    \includegraphics[width=0.8\linewidth]{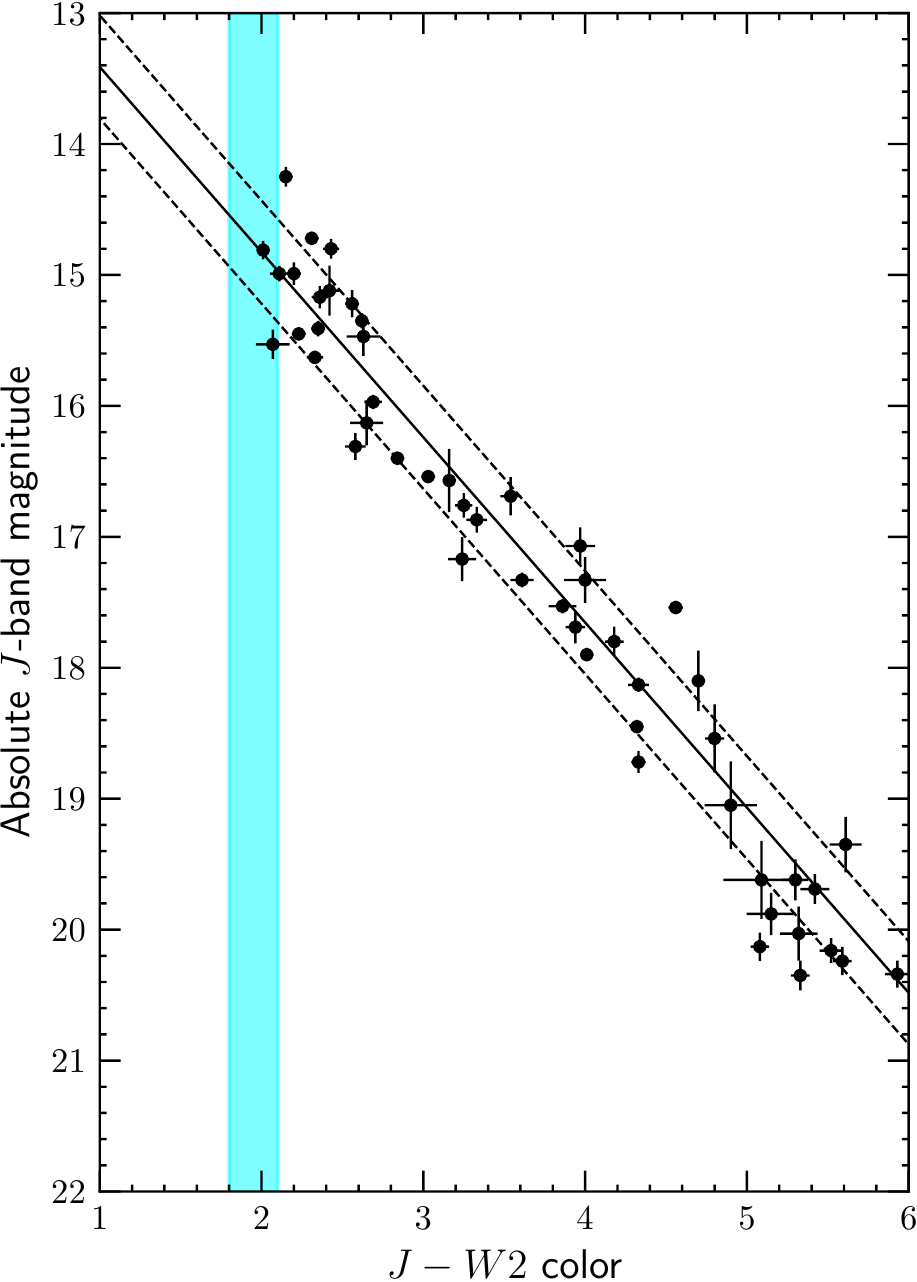}
    \caption{NIR color-magnitude diagram for cold brown dwarfs (class T and Y). Black points are taken from the homogenized dataset on late T-dwarfs and Y dwarfs presented by \citet{leggett2017}. The solid black line is a linear fit to the points. The dashed black lines are parallel to the solid line, and are offset by one standard deviation between the fit and the data points. The cyan shaded region shows the constraint on \jjj's NIR colors from Table\,\ref{tab:phot-astro}}
    \label{fig:color-mag}
\end{figure}

\subsection{NIR astrometry and proper motion}
The GNIRS keyhole images only had the target and Star A detected within the field of view. We used Star A to apply a global offset and used the nominal plate scale and keyhole position angle to determine the position of \jjj. The uncertainties on the plate scale and position angle are not well determined but are likely about $1\%$ and $0^\circ.01$ respectively (priv. comm. Siyi Xu). Based on this, we conservatively adopt an uncertainty of $0\arcsec.2$ in \jjj's position derived from GNIRS keyhole images.

The NIRI images allowed us to solve for offsets and distortions as many stars were detected. The exposures were set to the \texttt{USNO-B1} astrometric frame. A final offset correction on the extracted position of \jjj\, was applied such that the median offset of field stars in the {\em gaia} DR2 catalog was zero. We checked the \texttt{sextractor} extracted positions of other in-field stars that were comparable in brightness to Star\,A and found the astrometric accuracy to be about $0\arcsec.2$ which is likely dominated by uncertainties in our solution for plate scale and distortion terms. We note that the NIRI and GNIRS positions agree within errors.

\begin{figure}
    \centering
    \includegraphics[width=0.6\linewidth]{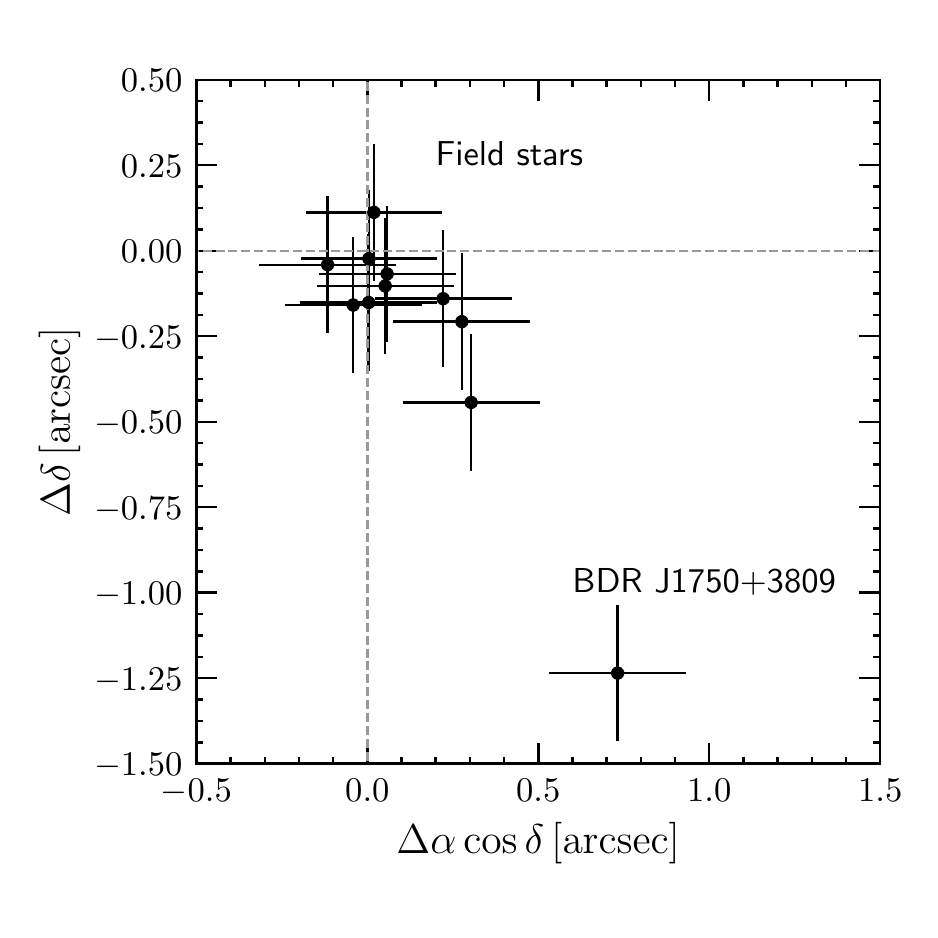}
    \caption{Astrometric offset of field stars and \jjj\, between the UKIRT and NIRI exposures, separated by about six years.}
    \label{fig:pm}
\end{figure}

We determined the proper motion of \jjj,\, using the UKIRT and NIRI \ch exposures because the NIRI $H$-band exposure had worse seeing. Fig. \ref{fig:pm} shows the offset of field stars and \jjj\, between the two images.

The apparent proper motion between the UKIRT exposure and the NIRI exposures is $-120\,{\rm mas/yr}$, and $200\,{\rm mas/yr}$ along the RA and DEC axes respectively. Based on our astrometric accuracy, we estimate the error in these estimates to be about $30\,{\rm mas/yr}$. We do not have sufficient number of measurements to simultaneously solve for parallax, proper motion and any orbital shift due to binarity. 
\end{document}